\documentclass{aa}

\usepackage{color}
\usepackage{graphicx}
\usepackage{natbib,twoopt}
\usepackage[varg]{txfonts}
\usepackage{gensymb}
\usepackage{lscape}

\bibpunct{(}{)}{;}{a}{}{,}

\makeatletter\renewcommand\theenumi{\@alph\c@enumi}\makeatother

\def\hi{\relax \ifmmode {\mbox H\,\texjtsc{i}}\else H\,{\scshape i}\fi}
\def\hii{\relax \ifmmode {\mbox H\,\textsc{ii}}\else H\,{\scshape ii}\fi}\def\nii{\relax \ifmmode {\mbox N\,\textsc{ii}}\else N\,{\scshape ii}\fi}\def\oi{\relax \ifmmode {\mbox O\,\textsc{i}}\else O\,{\scshape i}\fi}
\def\oii{\relax \ifmmode {\mbox O\,\textsc{ii}}\else O\,{\scshape ii}\fi}\def\oiii{\relax \ifmmode {\mbox O\,\textsc{iii}}\else O\,{\scshape iii}\fi}
\def\cii{\relax \ifmmode {\mbox C\,\textsc{ii}}\else C\,{\scshape ii}\fi}\def\ciii{\relax \ifmmode {\mbox C\,\textsc{iii}}\else C\,{\scshape iii}\fi}
\def\civ{\relax \ifmmode {\mbox C\,\textsc{iv}}\else C\,{\scshape iv}\fi}\def\hei{\relax \ifmmode {\mbox He\,\textsc{i}}\else He\,{\scshape i}\fi}\def\heii{\relax \ifmmode {\mbox He\,\textsc{ii}}\else He\,{\scshape ii}\fi}
\def\mgii{\relax \ifmmode {\mbox Mg\,\textsc{ii}}\else Mg\,{\scshape ii}\fi}
\def\sii{\relax \ifmmode {\mbox S\,\textsc{ii}}\else S\,{\scshape ii}\fi}\def\neiii{\relax \ifmmode {\mbox Ne\,\textsc{iii}}\else Ne\,{\scshape iii}\fi}
\def\ariv{\relax \ifmmode {\mbox Ar\,\textsc{iv}}\else Ar\,{\scshape iv}\fi}
\def\ni{\relax \ifmmode {\mbox N\,\textsc{i}}\else N\,{\scshape i}\fi}
\def\ariii{\relax \ifmmode {\mbox Ar\,\textsc{iii}}\else Ar\,{\scshape iii}\fi}
\def\caii{\relax \ifmmode {\mbox Ca\,\textsc{ii}}\else Ca\,{\scshape ii}\fi}

\begin{document}


   \title{Shape of the oxygen abundance profiles \\in CALIFA face-on spiral galaxies}
   \titlerunning{Shape of the oxygen abundance profiles in CALIFA face-on spiral galaxies}
   
   \author{L.~S\'anchez-Menguiano\inst{1,2}\and S.~F.~S\'anchez\inst{3}\and I.~P\'erez\inst{2}\and R.~Garc\'ia-Benito\inst{1}\and B.~Husemann\inst{4}\and D.~Mast\inst{5,6}\and A.~Mendoza\inst{1}\and T.~Ruiz-Lara\inst{2}\and Y.~Ascasibar\inst{7,8}\and J.~Bland-Hawthorn\inst{9}\and O.~Cavichia\inst{10}\and A.~I.~D\'iaz\inst{7,8}\and E.~Florido\inst{2}\and L.~Galbany\inst{11,12}\and R.~M.~G\'onzalez~Delgado\inst{1}\and C.~Kehrig\inst{1}\and R.~A.~Marino\inst{13,14}\and I.~M\'arquez\inst{1}\and J.~Masegosa\inst{1}\and J.~M\'endez-Abreu\inst{15}\and M.~Moll\'a\inst{16}\and A.~del~Olmo\inst{1}\and E.~P\'erez\inst{1}\and P.~S\'anchez-Bl\'azquez\inst{7,8}\and V.~Stanishev\inst{17}\and C.~J.~Walcher\inst{18}\and \'A.~R.~L\'opez-S\'anchez\inst{19,20}\and the CALIFA collaboration}
   \authorrunning{L.~S\'anchez-Menguiano et al.}

   \institute{Instituto de Astrof\'isica de Andaluc\'ia (CSIC), Glorieta de la Astronom\'ia s/n, Aptdo. 3004, E-18080 Granada, Spain\\
              \email{lsanchez@iaa.es}
         \and Dpto. de F\'isica Te\'orica y del Cosmos, Universidad de Granada, Facultad de Ciencias (Edificio Mecenas), E-18071 Granada, Spain
         \and Instituto de Astronom\'ia, Universidad Nacional Aut\'onoma de M\'exico, A.P. 70-264, 04510, M\'exico, D.F.
         \and European Southern Observatory (ESO), Karl-Schwarzschild-Str. 2, 85748 Garching b. M\"unchen, Germany
         \and Instituto de Cosmologia, Relatividade e Astrof\'isica - ICRA, Centro Brasileiro de Pesquisas F\'isicas, Rua Dr. Xavier Sigaud 150, CEP 22290-180, Rio de Janeiro, RJ, Brazil
         \and Observatorio Astron\'omico de C\'ordoba, Universidad Nacional de C\'ordoba, Argentina
         \and Departamento de F\'isica Te\'orica, Universidad Aut\'onoma de Madrid, Cantoblanco, E28049, Spain
         \and Astro-UAM, UAM, Unidad Asociada CSIC
         \and Sydney Institute for Astronomy, School of Physics A28, University of Sydney, NSW 2006, Australia
         \and Instituto de F\'isica e Qu\'imica, Universidade Federal de Itajub\'a, Av. BPS, 1303, 37500-903, Itajub\'a-MG, Brazil
         \and Millennium Institute of Astrophysics MAS, Nuncio Monse\~nor S\'otero Sanz 100, Providencia, 7500011 Santiago, Chile
         \and Departamento de Astronom\'ia, Universidad de Chile, Camino El Observatorio 1515, Las Condes, Santiago, Chile
         \and CEI Campus Moncloa, UCM-UPM, Departamento de Astrof\'isica y CC. de la Atm\'osfera, Facultad de CC. F\'isicas, Universidad Complutense de Madrid, Avda. Complutense s/n, 28040 Madrid, Spain
         \and Department of Physics, Institute for Astronomy, ETH Z\"urich, CH-8093 Z\"urich, Switzerland
         \and School of Physics and Astronomy, University of St Andrews, SUPA, North Haugh, KY16 9SS St Andrews, UK
         \and CIEMAT, Avda. Complutense 40, E-28040 Madrid, Spain
         \and Department of Physics, Chemistry and Biology, IFM, Link\"oping University, SE-581 83 Link\"oping, Sweden
         \and Leibniz-Institut f\"ur Astrophysik Potsdam (AIP), An der Sternwarte 16, D-14482 Potsdam, Germany
         \and Australian Astronomical Observatory, PO Box 915, North Ryde, NSW 1670, Australia
         \and Department of Physics and Astronomy, Macquarie University, NSW 2109, Australia\\
             } 

   \date{Received 25 September 2015 / Accepted 12 December 2015}


\abstract{We measured the gas abundance profiles in a sample of 122 face-on spiral galaxies observed by the CALIFA survey and included all spaxels whose line emission was consistent with star formation. This type of analysis allowed us to improve the statistics with respect to previous studies, and to properly estimate the oxygen distribution across the entire disc to a
distance of up to 3-4 disc effective radii (r$_e$). We confirm the results obtained from classical \hii\, region analysis. In addition to the general negative gradient, an outer flattening can be observed in the oxygen abundance radial profile. An inner drop is also found in some cases. There is a common abundance gradient between 0.5 and 2.0 r$_e$ of \mbox{$\alpha_{O/H} = -\,0.075\,\rm{dex}/r_e$} with a scatter of \mbox{$\sigma = 0.016\,\rm{dex}/r_e$} when normalising the distances to the disc effective radius. By performing a set of Kolmogorov-Smirnov tests, we determined that this slope is independent of other galaxy properties, such as morphology, absolute magnitude, and the presence or absence of bars. In particular, barred galaxies do not seem to display shallower gradients, as predicted by numerical simulations. Interestingly, we find that most of the galaxies in the sample with reliable oxygen abundance values beyond $\sim 2$ effective radii (57 galaxies) present a flattening of the abundance gradient in these outer regions. This flattening is not associated with any morphological feature, which suggests that it is a common property of disc galaxies. Finally, we detect a drop or truncation of the abundance in the inner regions of 27 galaxies in the sample; this is only visible for the most massive galaxies.}


\maketitle

\section{Introduction}\label{sec:intro}

Understanding how disc galaxies form and evolve is one of the greatest challenges in galactic astronomy. Some of the remaining
unsolved fundamental questions are comprehending the processes that are involved in the assembly of galaxies of different masses, the relative importance of mergers versus continuous gas accretion infall into the disc, the rate of metal enrichment, and the angular momentum transfers during these processes.

The distribution of gas chemical abundances and stellar parameters as well as their variation in space and time are important tools for answering these questions on the evolution of discs in spiral galaxies. Infall models of galaxy formation predict that spiral discs build up through accretion of material, which leads to an inside-out growth \citep{matteucci1989,molla1996,boissier1999}. The accretion brings gas into the inner parts of the discs, where it reaches high densities that trigger violent and quite efficient star formation. Thus, there is a fast reprocessing of gas in the inner regions, which leads to a population of old, metal-rich stars surrounded by a high-metallicity gaseous environment, while the outer regions are populated by younger, metal-poor stars formed from poorly enriched material. 
The first evidence that supports this scenario for disc evolution comes from studies on stellar colour profiles in nearby galaxies, which find bluer colours in the outer parts \citep[e.g.][]{dejong1996,macarthur2004,taylor2005,munozmateos2007}. This blueing can be interpreted on the basis of a change in the disc scale-length as a function of the observed wavelength band.
This result is predicted by models based on the inside-out growth \citep{prantzos2000}. To explain the nature of these colour gradients, recent works have analysed the radial change in the star formation history \citep[SFH,][]{sanchezblazquez2009,perezjimenez2013}. 

Another independent result that is consistent with this scenario comes from the weak dependence of disc galaxies with redshift
on the stellar mass-size relation. According to the inside-out growth of discs, galaxies are expected to increase their scale
lengths with time as they grow in mass \citep{barden2005,trujillo2004,trujillo2006}, resulting in a constant mass-size relation with cosmic time.
 
In this context, the study of the interstellar medium (ISM) is crucial to understand the chemical evolution of galaxies, since the enriched material is expelled during the evolution of stars, is mixed with the already existing gas, and condenses to form new stars. The ISM is fundamentally gaseous, and its chemical abundance can be derived by analysing spectroscopic features, that is, nebular emission lines. These emission lines are excited by photoionisation of the interstellar gas by hot and young massive stars \citep{aller1984,osterbrock1989}, which form clouds of ionised hydrogen (\hii\, regions) where star formation (among other processes) takes place. As oxygen is the most abundant heavy element in Universe, this makes it the best proxy for the total gas metallicity. 

The study of the Milky Way (MW) is also an exceptional tool for our understanding of galaxy evolution, mainly because we
can observe both the stellar and the gaseous components in greater detail than in other galaxies. Among other chemical properties of our Galaxy, the gas abundance gradient has been extensively studied \citep[e.g.][]{shaver1983, deharveng2000, pilyugin2003, esteban2005, quireza2006, rudolph2006, balser2011}; it is still not properly traced, however, especially in the inner parts. Therefore, complementary information coming from data of external galaxies would help us to overcome this problem.

The study of the gas metallicity in external individual galaxies using spectroscopic data allows us to shed light on fundamental physical properties of galaxies, such as SFR \citep[e.g.][]{ellison2008,laralopez2010,lopezsanchez2010}, mass and luminosity \citep[e.g.][]{lequeux1979,skillman1989,tremonti2004,rosalesortega2012}, effective yield and rotation velocity \citep[e.g.][]{vilacostas1992,garnett2002,pilyugin2004,dalcanton2007}, or stellar-to-gas fraction \citep[e.g.][]{zahid2014,ascasibar2015}. Moreover, studying these relations at different redshifts can help us to understand the assembly history and evolution of galaxies \citep[e.g.][]{kobulnicky2000,maiolino2008,mannucci2009,mannucci2010,belli2013}. 
The inside-out scenario is not only supported by studies focused on the stellar content in galaxies. Gas metallicity studies have also been key elements in favour of such disc growth, predicting a relatively quick self-enrichment with oxygen abundances and an almost universal negative metallicity gradient once it is normalised to the galaxy optical size \citep{boissier1999,boissier2000}. Several observational studies have found this radial decrease in the oxygen abundance along the discs of nearby galaxies \citep[e.g.][]{vilacostas1992,zaritsky1994,vanzee1998,bresolin2009,moustakas2010,rich2012}.

However, gas metallicity studies have also presented evidence of the existence of some behaviours in the oxygen abundance profiles that deviate from the pure inside-out scenario: A decrease or a nearly flat distribution of the abundance in the innermost region of discs, first observed by \citet{belley1992}; and a flattening in the gradient in the outer regions measured in several works \citep[][among others]{martin1995,vilchez1996,roy1997}. These features have been theoretically suggested to be motivated, for instance, by the presence of radial migration \citep{minchev2011, minchev2012}. Nevertheless, their origin is still unknown.

All these spectroscopic studies were limited by statistics, either in the number of observed \hii\, regions or in the coverage of these regions across the galaxy surface. The advent of integral field spectroscopy (IFS) techniques offers astronomers the opportunity to overcome these limitations by tracing the distribution of ionised gas and estimating spatially resolved chemical abundances for the gas phase. Its two-dimensional spatial coverage allows us to extract several hundreds or even thousands of spectra across the entire galaxy extent. This enables studying the variation of gas properties throughout the whole disc. 

Moreover, IFS surveys offer the opportunity of extending the study to a large number of objects, allowing for meaningful statistical analysis. However, until recently, this technique was rarely used in a survey mode. There were only a few exceptions such as the SAURON survey \citep{bacon2001} and the Disk Mass Survey \citep{bershady2010}.

These pioneering projects were not optimal for a statistical study of the properties of \hii\, regions because they incompletely covered the full extent of the galaxies, among other reasons. Such a statistical study started with the development of the PINGS project \citep{rosalesortega2010}, which acquired IFS mosaic data for a dozen very nearby galaxies. This project was followed by the observation of a larger sample of face-on spiral galaxies \citep{marmolqueralto2011} as part of the feasibility studies for the CALIFA survey \citep{sanchez2012a}. The advent of CALIFA allowed extending the study to much more representative samples of nearby galaxies by covering all morphologies.

Based on large-statistics samples of \hii\, regions extracted from galaxies observed by these programs, \citet{sanchez2012b,sanchez2014} studied the distribution of metals within star-forming galaxies and provided the strongest evidence so far for a characteristic gas abundance gradient out to two effective radii ($r_e$). These studies also confirmed the behaviours mentioned above that deviate from this gradient, as previously observed by other IFS works on individual galaxies \citep[e.g.][]{bresolin2009,sanchez2011,rosalesortega2011,bresolin2012,marino2012}. However, by selecting \hii\, regions they did not take advantage of the full capability of an IFS study and restricted the study to isolated areas of the galaxies. 

Current IFS surveys (e.g. ATLAS$^{\rm 3D}$, \citealt{cappellari2011}; CALIFA, \citealt{sanchez2012a}; SAMI, \citealt{croom2012}; MaNGA, \citealt{bundy2015}) have shown the potential of this kind of data to deliver important insights on this and other key questions about the formation and evolution of galaxies at low redshifts. Recent articles have proved the power of this tool to properly map the spatially resolved properties of galaxies by using the full two-dimensional (2D) information spaxel by spaxel that is provided by these surveys \citep[e.g.][]{papaderos2013, singh2013, davies2014, galbany2014, barrera2015, belfiore2015, gomes2015c, ho2015, holmes2015, li2015, martinnavarro2015, wilkinson2015}.

In this work, we make use of full 2D information in analysing CALIFA data spaxel by spaxel with the goal of characterising the radial gas abundance profile in a sample of face-on spiral galaxies. We not only focus on the broadly analysed gradient of these profiles, but also study other features that deviate from the simple negative trend, such as inner drops and outer flattenings. We also aim to compare the results with those obtained following the classical procedure of analysing \hii\, regions. A spaxel-by-spaxel study allows us to improve the statistics with respect to previous studies on the topic and also offers the possibility of properly estimating the oxygen distribution across the entire discs over a distance of up to 3-4 disc effective radii. A proper 2D study of the oxygen abundance distribution that analyses possible azimuthal variations will be presented in a forthcoming work.

The structure of the paper is as follows. Section~\ref{sec:sample} provides a description of the sample and data we use in this study. Section~\ref{sec:analysis} describes the analysis required to extract the spaxel-wise information. We explain the procedure to detect the \hii\, regions analysed for comparison (Sect.~\ref{sec:regions}) and derive the corresponding oxygen abundance values using both methods (Sect.~\ref{sec:measurements}). Our results are shown in Sect.~\ref{sec:results}, where we study the oxygen abundance slope distribution (Sect.~\ref{sec:hist}), its dependence on different properties of the galaxies (Sect.~\ref{sec:types}), and the existence of a common abundance profile (Sect.~\ref{sec:gradient}). Finally, the discussion of the results and the main conclusions are given in Sect.~\ref{sec:discussion}.


\section{Data and galaxy sample}\label{sec:sample}

The analysed data were selected from the 939 galaxies that comprise the CALIFA mother sample \citep{sanchez2012a}. These galaxies were observed using the Potsdam Multi Aperture Spectrograph \citep[PMAS;][]{roth2005} at the 3.5m telescope of the Calar Alto observatory with a configuration called PPAK \citep{kelz2006}. This mode consists of 382 fibres of 2.7 arcsec diameter each, 331 of them (the science fibres) covering an hexagonal field of view (FoV) of $74''\rm{x}\, 64''$. To achieve a filling factor of $100 \%$ along the full FoV and increase the spatial resolution, a dithering scheme of three pointings was adopted. Two different setups were chosen for the observations: V500, with a nominal resolution of $\lambda/\Delta\lambda \sim 850$ at 5000 \AA\, (FWHM $\sim 6\,$\AA) and a wavelength range from 3745 to 7500 \AA, and V1200, with a better spectral resolution of $\lambda/\Delta\lambda \sim 1650$ at 4500 \AA\, (FWHM $\sim 2.7\,$\AA) and ranging from 3650 to 4840 \AA. The data analysed here were calibrated with version 1.5 of the reduction pipeline. More detailed information about the CALIFA sample, observational strategy and data reduction can be found in \citet{sanchez2012a}, \citet{husemann2013}, and \citet{garciabenito2015}. 

After following the standard steps for fibre-based IFS data reduction, the pipeline provides a regular-grid datacube, with $x$ and $y$ coordinates indicating the right ascension and declination of the target and $z$ being the step in wavelength for all galaxies in the sample. An inverse-distance weighted image reconstruction scheme was performed as interpolation method to reconstruct the datacube. As a result, we have individual spectra for each sampled spaxel of $1'' \times 1''$ and a final spatial resolution for the datacubes of FWHM $\sim 2.5$ arcsec. 

\vspace{1cm}
The subset of galaxies used in this work  was selected by adopting the following criteria:
\begin{enumerate}
\item Spiral galaxies with morphological types between Sa and Sm, including barred galaxies.
\item Face-on galaxies, with $i < 60 \degree$, to avoid uncertainties induced by inclination effects.
\item Galaxies with no evident signatures of interaction or merging (i.e. tails, bridges, rings, etc.).
\item Galaxies with H$\alpha$ detected along different galactocentric distances with a signal-to-noise ratio (S/N) for the spaxels above 4 on average.
\end{enumerate}
The classification according to morphological type and into interacting or non-interacting galaxy was based on the visual inspection carried out by \citet[][see details in the article]{walcher2014}. After imposing these restrictions, the galaxy sample was reduced to 204 galaxies. From these, we only analysed the 129 galaxies that have finally been observed by the CALIFA collaboration with the V500 setup.

\begin{figure*}
\resizebox{\hsize}{!}{\includegraphics{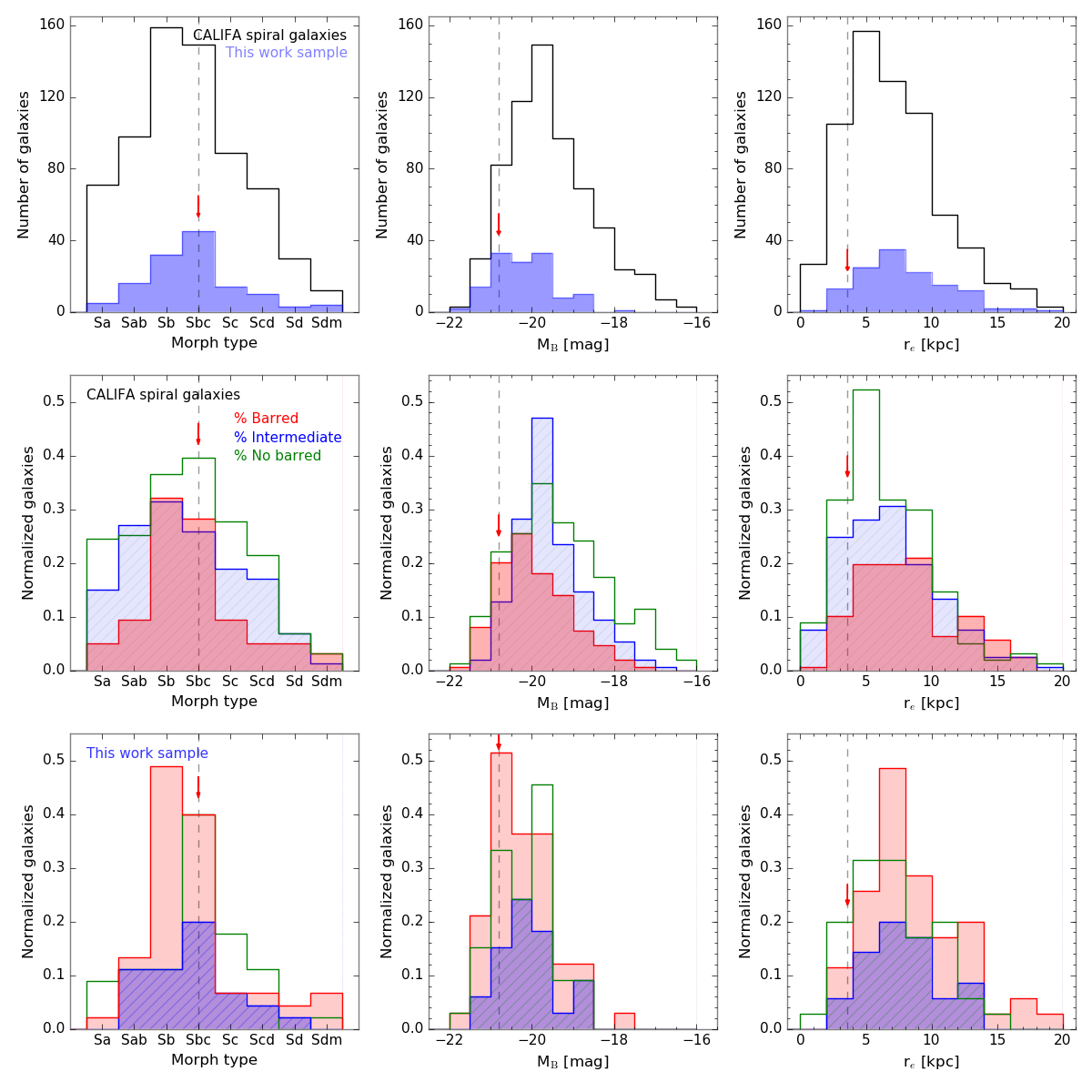}}
\caption{Distribution of morphological types (top left), absolute $B$-band magnitudes (top middle), and disc effective radii (top right) of the spiral galaxies in the CALIFA mother sample (unfilled black histograms) and the galaxies selected in this work (filled blue histograms). The middle and bottom panels show the normalised distributions separated according to the presence or absence of bars: barred galaxies (red histograms), unbarred galaxies (green histograms) and not clearly identified galaxies (blue histograms). The dashed grey line and the red arrow mark the location of the MW in each panel. The values for the absolute $B$-band magnitude (M$_{\rm B}=-20.8$ mag) and the disc effective radius (r$_e=3.6$ kpc) are taken from \citet{karachentsev2004} and \citet[][considering $r_e \sim 1.67 r_d$]{bovy2013} respectively.}
\label{fig:histograms}
\end{figure*}

Figure~\ref{fig:histograms} shows the comparison of the distribution of morphological types, absolute $B$-band magnitudes, and disc effective radii between the spiral galaxies in the CALIFA mother sample and the sample used in this study. There is a clear deficiency of earlier (Sa-Sab) and later (Sc-Sdm) spirals; the sample is dominated by intermediate galaxies. This may be due to the imposition of ionised gas throughout the discs, which we have prioritised to perform a detailed 2D study of the gas metallicity. The distribution of galaxies according to their absolute magnitude clearly shows an absence of faint galaxies, with values above -18 mag. This fact is a consequence of a selection effect in the definition of the CALIFA mother sample \citep{walcher2014}. The CALIFA mother sample was created by applying a size selection criterion defined by a minimum apparent isophotal size. A size-limited sample like
this favours inclined over face-on sytems because the inclination increases the apparent isophotal size (because the surface brightness increases). This effect causes these inclined galaxies to dominate the low-luminosity population of galaxies. Because we selected only face-on galaxies with $i< 60 \degree$, we automatically discarded all these faint galaxies. On the other hand, a correlation between the morphological type of the galaxies and the mass \citep[thus, with the luminosity, see e.g.][]{roberts1994,gonzalezdelgado2015} has been found, where later types present lower masses (and luminosities). This contributes to the deficiency of later spirals that is found in the sample because of the lack of low-luminosity galaxies and the correlation between these two parameters. For the distribution of galaxies according to their disc effective radius, we finally
obtained similar distributions for the CALIFA spiral galaxies and our sample: the sample is dominated by galaxies with r$_e$ between 4 and 10 kpc. Furthermore, there are no significant differences in the distribution when the galaxies are separated into barred and non-barred galaxies, galaxies of all sizes are present in both cases.

\begin{figure}
\resizebox{\hsize}{!}{\includegraphics{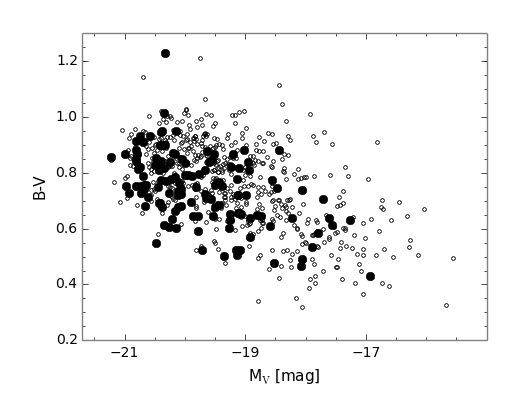}}
\caption{Distribution of the spiral galaxies in the CALIFA mother sample (empty small dots) and the galaxies selected in this work (filled large dots) in the $\left(B-V\right)$ vs $M_V$ colour-magnitude diagram.}
\label{fig:colormag}
\end{figure}

Figure~\ref{fig:colormag} shows the distribution of our sample (filled dots) and the total sample of CALIFA spiral galaxies (empty dots) along the $\left(B-V\right)$ vs $M_V$ colour-magnitude diagram. Our sample covers the same range as the CALIFA spirals above $M_V \sim -17$ mag, excluding the faint galaxies below this limit for the reasons explained above.

We note that with the limitations we mentioned, which are linked to the criteria we adopted to define the sample, the selected set of galaxies is well suitable to carry out the study presented here, that is, a detailed 2D study of the radial gas metallicity distribution in spiral galaxies.


\section{Analysis}\label{sec:analysis}

The main goal of this study is to characterise the radial abundance profiles in the galaxy sample using the full 2D information and compare it with the results obtained using only the \hii\, regions. In this section we describe the procedure followed to select the spaxels, analyse their individual spectra, and to derive the corresponding oxygen abundance. We also explain how we have detected the \hii\, regions used for comparison and the subsequent analysis.

\subsection{Measurement of the emission lines}\label{sec:analysis1}

In the spectrum of a galaxy (or a portion of it), the emission lines are superimposed on the underlying stellar spectrum. To accurately measure the emission line fluxes, the stellar contribution must be estimated and subtracted from the galaxy spectrum to derive a pure gas spectrum (allowing for the contribution of noise from the stellar populations) for each individual spaxel (or \hii\, region). 

Several tools have been developed to model the underlying stellar population and decouple it from the emission lines \citep[e.g.][]{cappellari2004, cidfernandes2005, ocvirk2006, sarzi2006, koleva2009, sanchez2011}. Most of them are based on the assumption that the star formation history (SFH) of a galaxy can be approximated as a sum of discrete star formation bursts and, therefore, that the stellar spectrum can be considered as the result of the combination of spectra of different simple stellar populations (SSP) with different ages and metallicities.
 
In this work, we made use of a fitting package named FIT3D\footnote{\url{http://www.astroscu.unam.mx/~sfsanchez/FIT3D}} to model both the continuum emission and the emission lines. This tool uses an SSP template grid that comprises 156 individual populations covering 39 stellar ages between 0.001 and 14.1 Gyr and four metallicities between 0.004 and 0.03. This grid combines the Granada models from \citet{gonzalezdelgado2005} for $t < 63$ Myr with those provided by the MILES project \citep{vazdekis2010,falconbarroso2011} for older ages \citep[following][]{cidfernandes2013}. This way, FIT3D fits each spectrum by a linear combination of the SSP templates that are collected in the library after correcting for the appropriate systemic velocity and velocity dispersion (including the instrumental dispersion, which dominates the total observed dispersion) and taking into account the effects of dust attenuation. We adopted the \citet{cardelli1989} law for the stellar dust extinction with $R_{V} = 3.1$. 

To measure the emission line fluxes and after the stellar component is subtracted, FIT3D performs a multi-component fitting using a single Gaussian function per emission line plus a low-order polynomial function. When more than one emission line was fitted simultaneously (e.g. for doublets and triplets like the [\nii] lines), the systemic velocity and velocity dispersion were forced to be equal to decrease the number of free parameters and increase the accuracy of the deblending process. The measured line fluxes include all lines required in determining the gas metallicity using strong-line methods, that is, H$\alpha$, H$\beta$, \mbox{[\oii]~$\lambda3727$}, \mbox{[\oiii]~$\lambda4959$}, \mbox{[\oiii]~$\lambda5007$}, \mbox{[\nii]~$\lambda6548$}, \mbox{[\nii]~$\lambda6584$}, \mbox{[\sii]~$\lambda6717,$} and \mbox{[\sii]~$\lambda6731$}. FIT3D provides the intensity, equivalent width (EW), systemic velocity, and velocity dispersion for each emission line. The statistical uncertainties in the measurements were calculated by propagating the error associated with the multi-component fitting and taking into account the S/N at the spectral region.

As indicated above, FIT3D fits both the underlying stellar population and the emission lines. In addition to the parameters derived for the emission lines, the fitting algorithm therefore provides information related to the stellar population: the luminosity-weighted ages and metallicities, the average dust attenuation, the mass-weighted ages and metallicities, the average mass-to-light ratio, and the individual weights of the multi-SSP decomposition that in essence trace the SFH.

The entire procedure of fitting and subtracting the underlying stellar population and measuring the emission lines using FIT3D is described in more detail in \citet{sanchez2011} and \citet{sanchez2015b}.

We note that all these parameters (both stellar and gas) were derived spaxel by spaxel for each individual spectrum of the datacubes, providing the sets of $2$D maps that are the base of our analysis.

\subsection{Extracting information spaxel by spaxel}\label{sec:analysis2}
As a result of the FIT3D fitting process, we obtained the set of $2$D intensity maps for the emission lines that are required to determine the gas metallicity. To guarantee realistic measurements of the emission line fluxes for each spaxel, we adopted a lower limit below which we considered that the fluxes are of the same order as the continuum error. In this way, we discarded the spaxels whose emission line fluxes employed in the determination of the oxygen abundance are lower than $1\,\sigma$ over the continuum level. From all the spaxels with flux values above this limit, we now selected those that are associated with star formation (SF). 

The intensities of strong lines were broadly used to discern between different types of emission according to their main excitation source (i.e. starburst or AGN) throughout the so-called diagnostic diagrams \citep[e.g.][]{baldwin1981,veilleux1987}. In most cases these diagrams are very useful in distinguishing between strong ionisation sources, such as classical \hii\,/SF regions and powerful AGNs. However, they are less accurate in distinguishing between low-ionisation sources, such as weak AGNs, shocks, and/or post-AGBs stars \citep{stasinska2008,cidfernandes2011}. Alternative methods based on a combination of the classical line ratios and additional information regarding the underlying stellar population have been proposed, for instance, the so-called WHAN diagram \citep{cidfernandes2011}. This diagram uses the EW(H$\alpha$) to take into account weak AGNs and `retired' galaxies, that is, galaxies that have stopped forming stars and are ionised by hot low-mass evolved stars.

The most commonly used diagnostic diagram was proposed by \citet[][hereafter BPT diagram]{baldwin1981}. This diagram makes use of the \mbox{[\nii]~$\lambda6584$/H$\alpha$} and \mbox{[\oiii]~$\lambda5007$/H$\beta$} line ratios, which are less affected by dust attenuation because of their proximity in wavelength space. Different demarcation lines for BPT diagram have been proposed to distinguish between SF regions and AGNs. The most popular are the \citet{kauffmann2003} and \citet{kewley2001} curves. Pure \hii\,/SF regions are considered to be below the \citet{kauffmann2003} curve and AGNs above the \citet{kewley2001} curve. The area between the two curves is broadly and erroneously assigned to a mixture of different ionisation sources, since pure SF regions can also be found here. 

These two demarcation lines have a different origin. The \citet{kewley2001} curve was derived theoretically using photoionisation models and corresponds to the maximum envelope for ionisation produced by OB stars. The \citet{kauffmann2003} curve has an empirical origin, based on the analysis of the emission lines for the integrated spectra of SDSS galaxies. It describes the envelope for classical \hii\,/SF regions well that are found in the discs of late-type spiral galaxies. However, it excludes certain kinds of SF regions that have already been found above this demarcation line (\citealt{kennicutt1989b,ho1997} and, more recently, \citealt{sanchez2014}).
Selecting \hii\,/SF regions based on the \citet{kauffmann2003} curve may therefore bias our sample towards classical disc regions. Moreover, it does not guarantee that other sources of non-stellar ionisation are excluded that might populate this area, such as weak AGNs, shocks, and/or post-AGBs stars. We adopted the \citet{kewley2001} curve to exclude strong AGNs and an EW criterion to exclude weak AGNs and `retired' emission \citep{cidfernandes2011}. However, we were more restrictive in the EW range than \citet{cidfernandes2011} and established the limit in 6 \AA\, to also guarantee a better S/N of the emission lines for all spaxels. For a detailed analysis of the weak emission procedure spaxel by spaxel using CALIFA data, see \citet{gomes2015c}.

In addition to the light from the ionised SF regions, there is a background of diffuse nebular emission that extends over the whole disc of the galaxies and can blur contribution of the SF regions, which is the subject of our study. However, most of the diffuse ionised emission has been excluded by the $1\,\sigma$ limit imposed to the flux of the selected spaxels and the EW criterion, since this emission is dominated by the stellar continuum. For a comparison of the location in the BPT diagram for low-ionisation emission sources see, for instance, \citet{kehrig2012,papaderos2013}, and \citet{gomes2015c}.

\begin{figure*}
\resizebox{\hsize}{!}{\includegraphics{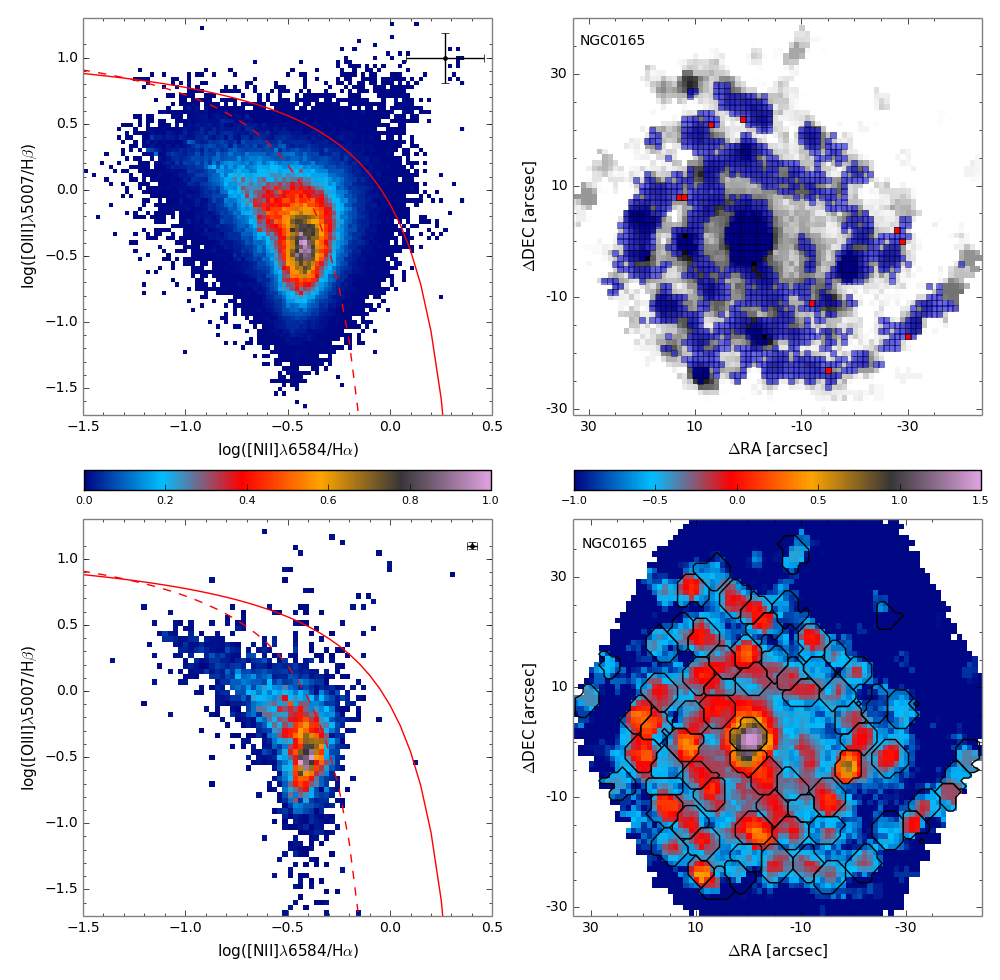}}
\caption{{\it Left panels:} Normalised density distribution of the spaxels with EW(H$\alpha$) above $6\,$\AA\, (top) and of the detected \hii\,/SF regions (bottom) along the BPT diagram. The solid and dashed lines in both panels represent the \citet{kewley2001} and \citet{kauffmann2003} demarcation curves. SF regions are considered to be below the \citet{kewley2001} curve. {\it Right panels:} Location of the spaxels classified as SF regions (blue dots) and AGNs (red dots) according to the BPT diagram superimposed on the IFS-based H$\alpha$ map derived for one galaxy of the sample, NGC~0165 (top) and a H$\alpha$ map in units of (log10) $10^{-16}$ erg s$^{-1}$ cm$^{-2}$ arcsec$^{-1}$ for NGC~0165, together with the detected \hii\, regions shown as black segmented contours (bottom).}
\label{fig:bpt_Hamaps}
\end{figure*}

The top left panel of Fig.~\ref{fig:bpt_Hamaps} shows the \mbox{[\oiii]~$\lambda5007$/H$\beta$} vs. \mbox{[\nii]~$\lambda6584$/H$\alpha$} diagnostic diagram for the spaxels in all 129 galaxies of our sample above the considered flux limit and with \mbox{EW(H$\alpha$)$\, > 6\,$\AA}. The solid and dashed lines represent the \citet{kewley2001} and \citet{kauffmann2003} demarcation curves, respectively. Some points dominated by SF ionisation might be present above the \citet{kewley2001} curve as a result of the errors of the considered emission lines. They are, therefore, excluded from further analysis by our criteria for selecting spaxels associated with SF activity. However, the spaxels that present larger errors are those with a low S/N, and they do not affect our conclusions significantly.

The top right panel of Fig.~\ref{fig:bpt_Hamaps} shows the location of the selected spaxels in a particular spiral galaxy of the sample, NGC~0165, over-plotted to the H$_\alpha$ map. Blue dots correspond to the spaxels classified as SF regions and red dots are those that lie higher than the \citet{kewley2001} curve and can therefore be associated with AGNs. The figure shows that the selected spaxels follow the H$_\alpha$ emission. The classification of red points as ionised by AGNs in this galaxy as well as in other cases is clearly false because of their distance to the galactic centres. This misclassification is most probably due to the errors in the considered emission lines. As in the previous case, they could cause some spaxels that are dominated by SF ionisation to lie higher than the \citet{kewley2001} curve in a similar way that errors could produce the opposite effect with spaxels associated to AGNs. As these spaxels represent only the $2\%$ for the whole sample, including them would not alter our results significantly, and thus they were discarded from the further analysis. For this galaxy, $1\,201$ spaxels were associated with SF regions. 

\subsection{Detection and selection of \hii\,regions}\label{sec:regions}

We detected the \hii\, regions and extracted the corresponding spectra using a semi-automatic procedure named HII{\tiny EXPLORER}\footnote{\url{http://www.astroscu.unam.mx/~sfsanchez/HII\_explorer}}. The procedure is based on two assumptions: (a) \hii\, regions are peaky and isolated structures with a strongly ionised gas emission, particularly H$\alpha$, that is significantly higher than the stellar continuum emission and the average ionised gas emission across the galaxy; (b) \hii\, regions have a typical physical size of about one  hundred or a few hundred parsecs \citep{gonzalezdelgado1997, oey2003, lopez2011}, which corresponds to a typical projected size of a few arcsec at the standard distance of the galaxies in the sample. 

A more detailed description of this algorithm can be found in \citet{sanchez2012b}, with a few modifications presented in \citet{sanchez2015a}. Basically, the main steps of the process are as follows: (i) First we create a narrow-band image of 120 \AA\, width centred on H$\alpha$ shifted at the redshift of each galaxy. (ii) This image is used as an input for HII{\tiny EXPLORER}. The algorithm detects the brightest pixel in the map and then adds all the adjacent pixels up to a distance of 3.5'' if their fluxes exceed 10\% of the peak intensity. After the first region is detected and separated, the corresponding area is masked from the input image and the procedure is repeated until no peak with a flux exceeding the median H$\alpha$ emission flux of the galaxy is found. The result is a segmentation FITS file describing the pixels associated with each detected \hii\, region. Finally, (iii) the integrated spectrum corresponding to each segmented region is extracted from the original datacube, and the corresponding position table of the detected area is provided.

After we extracted the spectra for the detected clumpy ionised regions, we applied the same analysis described in Sects.~\ref{sec:analysis1} and \ref{sec:analysis2}: each extracted spectrum was decontaminated by the underlying stellar population using FIT3D, and the emission line fluxes were measured by fitting each line with a Gaussian function. These line ratios were used to distinguish between the detected ionised regions, the ones associated with star formation. In a similar way as for individual spaxel spectra, using the BPT diagram \hii\,/SF regions were considered to be under the \citet{kewley2001} curve and present an \mbox{EW(H$\alpha$)$\, > 6$\AA}.

Figure~\ref{fig:bpt_Hamaps} (bottom left) shows the \mbox{[\oiii]~$\lambda5007$/H$\beta$} vs. \mbox{[\nii]~$\lambda6584$/H$\alpha$} diagnostic diagram for the \hii\,/SF regions. The solid and dashed lines represent the \citet{kewley2001} and \citet{kauffmann2003} demarcation curves, respectively. 

Figure~\ref{fig:bpt_Hamaps} (bottom right) shows an example of an H$\alpha$ map for one spiral galaxy of the sample, NGC~0165, where the location of the \hii\, regions is superimposed as black segmented contours. For this galaxy we detected 72 \hii\, regions.

\subsection{Measurement of the oxygen abundances}\label{sec:measurements}
A direct procedure to measure abundances from observed spectra requires using temperature-sensitive line ratios such as \mbox{[\oiii]~$\lambda\lambda4959,5007/[\oiii]~\lambda4363$}. This is known as the $T_e-$method \citep{peimbert1969,stasinska1978,pagel1992,vilchez1996,izotov2006}. However, some of these auroral or nebular lines are very faint, and they become even fainter as the metallicity increases (when a more efficient cooling mechanism begins to act through the metal lines, which produces a decrease in the temperature), and
eventually, they are too weak to be detected. Furthermore, as
a result of the weakness of the involved lines, the $T_e-$method can only be applied to nearby and low-metal objects for which very high S/N spectra are observable.

It is therefore necessary to look for indirect methods that allow us to estimate the chemical abundances. To do this, abundance indicators based on the relations between metallicity and the intensity of strong and more readily observable lines have been developed. These methods have first been proposed by \citet{alloin1979} and \citet{pagel1979}. Since then, several calibrators based on direct estimations of oxygen abundances \citep{zaritsky1994,pilyugin2000,denicolo2002,pettini2004,perezmontero2005,pilyugin2005,pilyugin2010,marino2013} and photoionisation models \citep{dopita1986,mcgaugh1991,kewley2002,kobulnicky2004,dopita2006,dopita2013,perezmontero2014} have been proposed and are widely used todays. For a revision of the different methods, their strengths and their caveats, see \citet{lopezsanchez2012}.

In this work we aim to derive the spatial distribution of the oxygen abundance across the considered galaxies. For this purpose, we used the emission line intensities derived spaxel by spaxel for the sample of \hii\,regions described before. We adopted the empirical calibrator based on the O3N2 index that was first introduced by \citet{alloin1979}:

\begin{equation}
{\rm O3N2} =  \log\left(\frac{[\oiii] \lambda5007}{{\rm H}\beta} \times \frac{{\rm H}\alpha}{[\nii] \lambda6584}\right)
.\end{equation}

This index (i) is only weakly affected by dust attenuation because of the close distance in wavelength between the lines in both ratios, (ii) presents a monotonic dependence on the abundance and (iii) uses emission lines covered by CALIFA wavelength range. One of the most popular calibrations for this index has been proposed by \citet[][hereafter PP04]{pettini2004}. However, this indicator lacks observational points at the high-metallicity regime \mbox{($12+\log\left({\rm O/H}\right) > 8.2$)} and instead
uses predictions from photoionisation models. Therefore, we here
adopted the improved calibration proposed by \citet[][hereafter M13]{marino2013}, where \mbox{$12+\log\left({\rm O/H}\right) = 8.533 - 0.214 \,\times\, {\rm O3N2}$}. This calibration uses $T_e$-based abundances of $\sim 600$ \hii\, regions from the literature together with new measurements from the CALIFA survey, providing the most accurate calibration to date for this index. The derived abundances have a calibration error of $\pm 0.08$ dex, and the typical errors associated with the pure propagation of the errors in the measured emission lines are about 0.05 dex.

\subsection{Oxygen abundance gradients}\label{sec:grad}

To derive the radial distribution of the oxygen abundance for each galaxy, we determined the position angle and ellipticity of the disc to obtain the deprojected galactocentric distances of the selected spaxels. These morphological parameters were derived by performing a growth curve analysis \citep[][hereafter S14]{sanchez2014}. The inclination was deduced by also assuming an intrinsic ellipticity for galaxies of $q=0.13$ \citep{giovanelli1994}:
\begin{equation}
\cos^2 i = \frac{(1-\epsilon)^2 - q^2}{1-q^2}
,\end{equation}
where $\epsilon$ is the ellipticity provided by the analysis and given by \mbox{$\epsilon=1-b/a$}, with $a$ and $b$ being the semi-major and semi-minor axes. We preferred not to correct for the inclination effects in galaxies with an inclination below $35 \degree$ because the uncertainties in the derived correction exceed the very small effect on the spatial distribution of the spaxels, even more when an intrinsic ellipticity is also considered.

We then derived the galactocentric distance for each spaxel, which was later normalised to the disc effective radius, as suggested in \citet{sanchez2012b,sanchez2013}. This parameter was derived from the disc scale-length based on an analysis of the azimuthal surface brightness profile (SBP), explained in \mbox{Appendix~A} of S14. Other normalisation length-scales were used for better comparison with other studies, such as the $r_{25}$ radius, which is defined as the radius corresponding to a surface brightness level of 25 mag/arcsec$^2$ in the SDSS $r$-band\footnote{Using the seventh data release \citep[DR7,][]{dr7}.}, and the physical scale of the galaxy, that is, the distances in kpc.

\begin{figure}
\resizebox{\hsize}{!}{\includegraphics{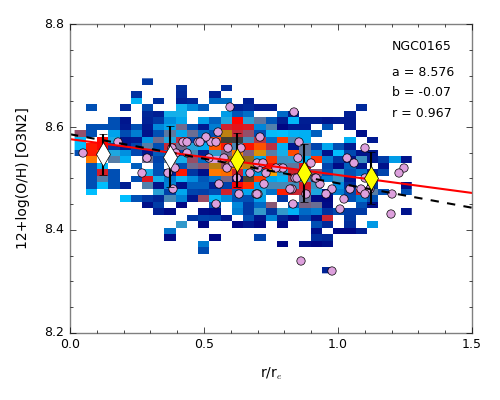}}
\caption{Radial density distribution of the spaxels in the oxygen abundance-galactocentric distance space for the same galaxy as in Fig.~\ref{fig:bpt_Hamaps} (right panels). The radial distances are deprojected and normalised to the disc effective radius. The diamonds represent the mean oxygen abundance values, with the error bars representing the corresponding standard deviations, for bins of 0.25 r$_e$ and the red solid line the error-weighted linear fit derived for values within the range between 0.5 and 2.0 r$_e$ (yellow diamonds). The parameters of the fit are shown in the upper right corner of the panels, including the zero point (a), the slope (b) and the correlation coefficient (r). The violet dots correspond to the oxygen abundances derived for the individual \hii\, regions, and the dashed black line is the linear regression for these points.}
\label{fig:grads_examples}
\end{figure}

Finally, we obtained the radial distribution of the oxygen abundance for each galaxy. To characterise this profile, we performed an error-weighted linear fit to the derived oxygen abundance mean values for radial bins of 0.25 $r_e$ within the range between 0.5 and 2.0 $r_e$. The radial binning was made to minimise possible azimuthal differences in the oxygen abundance distribution, and the size of the bins was chosen to match the seeing value. We excluded the innermost region \mbox{($r/r_e < 0.5$)}, which sometimes presents a nearly flat distribution or even a drop towards the centre \citep[e.g.][]{belley1992,rosalesortega2011,sanchez2012b,sanchez2014}. We also omitted the outer region \mbox{($r/r_e > 2.0$)}, which it is found to have a flattening in the abundance gradient for galaxies covering regions beyond $r_{25}$ \citep[e.g.][]{martin1995, vilchez1996, roy1997, vanzee1998, bresolin2009, rosalesortega2011, bresolin2012, marino2012,lopezsanchez2015}. The edges of the range were obtained based on a visual inspection of each individual galaxy, discarding the regions where we observed the mentioned features. The fitted interval has changed with respect to S14, simply because of a better space coverage that allowed us to refine the radial limits. The linear fit was weighted using the standard deviations of (mean) values within each bin and considered only the bins that contained at least nine values of the oxygen abundance. This minimum number of values required for each bin was determined to ensure a precision in the derived mean that
is ten times better than the dynamic range of abundance values covered in the fit, taking into account the dispersion in the measurements.

It is important to note that uncertainties in the determination of the deprojected galactocentric distances can significantly affect the derivation of the abundance gradients. Moreover, as the radial galactocentric distances are normalised to the disc effective radius, the uncertainties in the determination of the $r_e$ can also contribute to the scatter obtained on the final oxygen gradients. On one hand, performing Monte Carlo simulations, we obtained that an error of $5 \degree$ in the inclination and PA of the galaxies can produce a dispersion in the gradient distribution of at most 0.05 dex/$r_e$ (0.02 dex/$r_e$ on average). On the other hand, comparing different methods to derive the $r_e$ (as described in S14) and taking into account both the nominal errors and the differences among them, the overall contribution to the dispersion in the gradient distribution coming from the derivation of the $r_e$ is at most 0.04 dex/$r_e$ (0.01 dex/$r_e$ on average). All these uncertainties are well accounted for by our error estimation of the gradient (0.05 dex/$r_e$, see Sect.~\ref{sec:hist}).

In Fig.~\ref{fig:grads_examples} we present an example of the abundance gradient for the same galaxy shown in the right panels of Fig.~\ref{fig:bpt_Hamaps}, NGC~0165, using both the spaxel-wise information (colour map) and the individual \hii\, regions (violet dots). The diamonds represent the mean oxygen abundance values of the radial bins, with the error bars indicating the corresponding standard deviations. The red solid line is the error-weighted linear fit derived for values within the range between 0.5 and 2.0 r$_e$ (yellow diamonds), and the dashed black line is the linear regression for the individual \hii\, regions. This figure illustrates the procedure explained before to derive the oxygen abundance gradient. From the original 129 galaxies, we were able to fit 122 lines, and the remaining galaxies were discarded from further analysis because of the low number of spaxels associated with SF regions that are needed to carry out the linear fit. This final sample provides more than 185$\,$000 oxygen abundance values, $\sim 8\,230$ of them beyond two disc effective radii, and with more than 7$\,$100 \hii\, regions to compare with, $\sim 605$ beyond 2 $r_e$.


\section{Results}\label{sec:results}

With the procedure explained in the previous section we obtained the oxygen abundance gradient for the 122 galaxies in our sample. We describe the main properties of these abundance profiles below.

\subsection{Abundance gradient distribution}\label{sec:hist}

\begin{figure*}
\resizebox{\hsize}{!}{\includegraphics{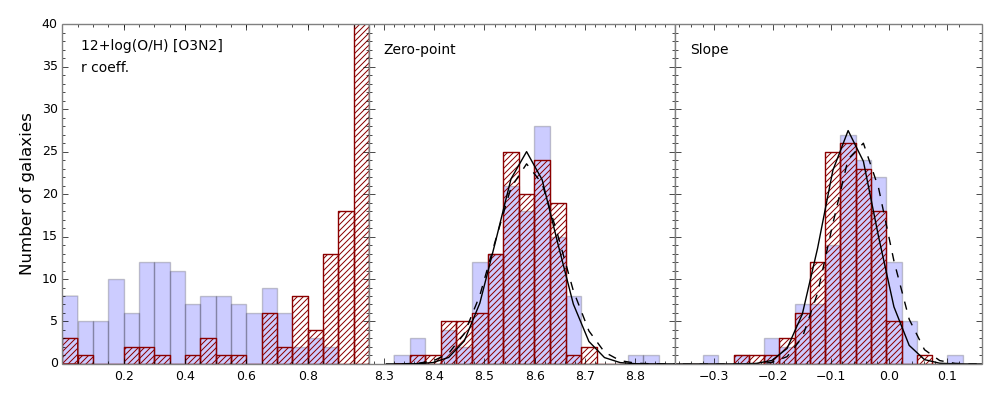}}
\caption{Distribution of correlation coefficients (left panel), zero points (middle panel), and slopes (right panel) of the linear fits derived for the oxygen abundance gradients of the final sample using spaxel-wise information (dashed red bars) and the individual \hii\, regions (filled blue bars). For both the zero point and slope distributions the lines represent the Gaussian distribution of the data (solid line for spaxels, dashed line for individual \hii\, regions), assuming the mean and standard-deviation of the distribution of each analysed parameter and sampled with the same bins.}
\label{fig:ox_hists}
\end{figure*}

Figure~\ref{fig:ox_hists} shows the distribution of the correlation coefficient, zero-point, and slope of the abundance gradients for the final sample using both the spaxel-wise information (red dashed histogram) and the individual \hii\, regions (blue filled histogram). 

We focus first on the spaxel-wise analysis. For almost all galaxies the correlation coefficient is larger than 0.75. If we perform a Student's t-test to check the significance of the correlation, we obtain that for $\sim 80\%$ of the galaxies the oxygen abundance and the radial distance (normalised to the disc effective radius) are significantly correlated to the $95\%$ level (0.05). 

The distribution of zero points ranges between 8.4 and 8.7, reflecting the mass-range covered by the sample as a consequence of the well-known $\mathcal{M}-\mathcal{Z}$ relation \citep[e.g.][]{tremonti2004,sanchez2013}. The presence of a peak in the distribution and a small standard deviation is the result of a bias in the selection of the sample, explained in Sect.~\ref{sec:sample}, which is due to a lack of low-luminosity galaxies. 

Finally, the distribution of slopes presents a characteristic value of $\alpha_{O/H} = -0.07 \,\rm{dex}/r_e$ with a standard deviation of $\sigma \sim 0.05 \,\rm{dex}/r_e$. We performed a Lilliefors test \citep{lilliefors1967} to assess the compatibility of the distribution with a Gaussian, obtaining a test statistics of 0.07 and a P-value of 0.62, showing that the distribution of slopes has a clear peak and is remarkably symmetric. We ran a Monte Carlo simulation to estimate the contribution of the errors in the derived slopes to the $\sigma$ of the distribution, obtaining that these errors can only explain  $49\%$ of the width distribution. We may have underestimated the errors involved in the determination of the slopes, particularly the effect of the inclination. Otherwise, the remaining $\sigma$ must have another origin that we investigate in more detail in Sect.~\ref{sec:types}. 

The analysis for the individual \hii\, regions leads us to very similar results. In this case, we have a wider distribution for the correlation coefficients, but we have to note that the number of points involved in the linear fit is larger using all the individual \hii\, regions, since we did not apply any kind of radial binning to the data. The correlation coefficient is larger than 0.32 for $\sim 60\%$ of the galaxies, which corresponds to a significance level of $98\%$ (0.02). The distribution of zero points covers almost the same range as for the spaxels (between 8.3 and 8.8), allowing us to draw the same conclusions. Finally, the distribution of slopes presents a characteristic value of $\alpha_{O/H} = -0.05 \,\rm{dex}/r_e$ with a standard deviation of $\sigma \sim 0.06 \,\rm{dex}/r_e$. The Lilliefors test gives a test statistic of 0.08 and a P-value of 0.28, very similar to the one described for the spaxel-wise analysis. The Monte Carlo simulation yields a contribution of $44\%$ of the errors in the derived slopes to the distribution width, again insufficient to explain the $\sigma$ of the distribution. 

If we use different scales to normalise the radial distances like $r_{25}$ and the physical scale of the galaxy (radius in kpc) for both the spaxel-wise and the individual \hii\, region analysis, we obtain in all cases a similar and narrow distribution, although for the physical scale the distribution is clearly asymmetric, with a tail towards large slopes. We note that our range of masses is narrow, and consequently, so is the range of $r_e$ and $r_{25}$, which in turn causes the distribution when normalising to the physical scale narrow as well, in contrast to what we should obtain for a wider range of masses. The different slope values are given in Table~\ref{tab:grads}.

\begin{table*}
\centering
\def\arraystretch{1.5}
\tabcolsep=0.17cm
   \caption{Oxygen abundance gradient slopes derived using different calibrators and distance normalisations.}
   \begin{tabular}{c@{\hspace{0.5cm}}ccc@{\hspace{0.7cm}}ccc}
      \hline
      {\large Calibrator}  & \multicolumn{3}{c}{{\large $\alpha_{\rm O/H}$}$\,-\,$spaxels} & \multicolumn{3}{c}{{\large $\alpha_{\rm O/H}$}$\,-\,$\hii\, regions}\\
        & [dex/$r_e$] & [dex/$r_{25}$] & [dex/kpc] & [dex/$r_e$] & [dex/$r_{25}$] & [dex/kpc]\\
      \hline
      O3N2 [M13] & $-0.07 \pm 0.05$ & $-0.08 \pm 0.06$ & $-0.009 \pm 0.008$ & $-0.05 \pm 0.06$ & $-0.06 \pm 0.06$ & $-0.008 \pm 0.010$\\
      O3N2 [PP04] & $-0.11 \pm 0.07$ & $-0.12 \pm 0.09$ & $-0.014 \pm 0.012$ & $-0.08 \pm 0.09$ & $-0.09 \pm 0.09$ & $-0.011 \pm 0.014$\\
      ONS [P10] & $-0.06 \pm 0.06$ & $-0.08 \pm 0.07$ & $-0.008 \pm 0.009$ & $-0.06 \pm 0.09$ & $-0.08 \pm 0.11$ & $-0.008 \pm 0.012$\\
      pyqz [D13] & $-0.14 \pm 0.09$ & $-0.15 \pm 0.10$ & $-0.019 \pm 0.014$ & $-0.11 \pm 0.09$ & $-0.13 \pm 0.11$ & $-0.015 \pm 0.015$\\
      \hline
   \end{tabular}
   \label{tab:grads}
\end{table*}

\subsubsection{Comparison with other calibrators}\label{sec:calibrators}

It is beyond the purpose of this work to make a detailed comparison of the oxygen abundance gradients derived using different empirical calibrators. However, we compare our results with those provided with other methods by deriving the oxygen radial distributions using some of the most commonly used empirical calibrators: (i) the technique proposed by PP04 for the O3N2 index, (ii) the \citet[][hereafter P10]{pilyugin2010} calibration for the ONS index, (iii) and the \citet[][hereafter D13]{dopita2013} calibration based on the MAPPINGS IV code developed by the authors. 

The PP04 calibration for the O3N2 index, as already mentioned in Sect.~\ref{sec:measurements}, is one of the most popular calibrations used for this index and is defined as $12+\log\left({\rm O/H}\right) = 8.73 - 0.32 \,\times\, {\rm O3N2}$. This calibration is not valid in the low-metallicity regime ($12+\log\left({\rm O/H}\right) < 8$), but as we do not reach this limit, this effect will not affect our results. The P10 ONS calibration makes uses of the $N_2/R_2$ and $S_2/R_2$ ratios (defined as $R_2 = [\oii] \left(\lambda3727 + \lambda3729\right)$, $N_2 = [\nii] \left(\lambda6548 + \lambda6584\right)$, $S_2 = [\sii] \left(\lambda6717 + \lambda6731\right)$) as temperature and metallicity indexes and is valid over the whole range of explored metallicities. The derived relations to determine the oxygen abundances using this calibration are given by P10. Finally, the D13 calibration is based on a grid of photoionisation models covering a wide range of abundance and ionisation parameters typical of \hii\, regions in galaxies. This calibration can be used through a Python module implemented by the authors, known as pyqz, which is publicly available\footnote{\url{http://dx.doi.org/10.4225/13/516366F6F24ED}}.

Table~\ref{tab:grads} shows a comparison among the oxygen abundance slopes derived using the different calibrators and the different normalisations for the radial distance described before. In this table we present the values using both the spaxels and the individual \hii\, regions.

\begin{figure}
\resizebox{\hsize}{!}{\includegraphics{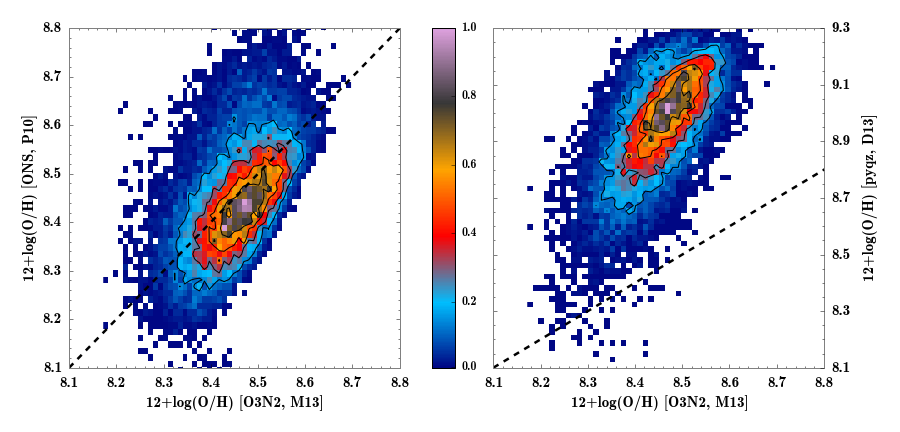}}
\caption{Comparison of the oxygen abundance distribution derived using the calibration proposed by M13 for the O3N2 index with the distribution derived using the P10 calibration for the ONS index (left panel) and the calibration based on pyqz code (D13, right panel). The black contours show the density distribution of the SF spaxels, the outermost one including 80\% of the total number of spaxels and decreasing 20\% in each consecutive contour. The black dashed lines indicate the 1:1 relation between the represented calibrators.}
\label{fig:ox_comp}
\end{figure}

We also illustrate this comparison in Fig.~\ref{fig:ox_comp}. The left panel represents the distribution of oxygen abundances derived using the M13 calibration for the O3N2 index vs the P10 calibration for the ONS index. In the right panel we show the same distribution of the M13 calibration, but this time vs the D13 calibration based on the pyqz code. Both panels show a tight correlation between the compared calibrators. This allows us to conclude that our qualitative results are not contingent upon the choice of the used calibrator, although the actual measured values for the abundance gradients may change. Similar conclusions were stated by \citet{ho2015} for a different sample of galaxies.

For the sake of clarity, below we only show the results of this article based on the use of the M13 calibrator. However, we have reproduced the analysis using all the proposed calibrators, with no significant differences between the obtained results.

\subsection{Abundance gradients by galaxy types}\label{sec:types}

\begin{figure*}
\resizebox{\hsize}{!}{\includegraphics{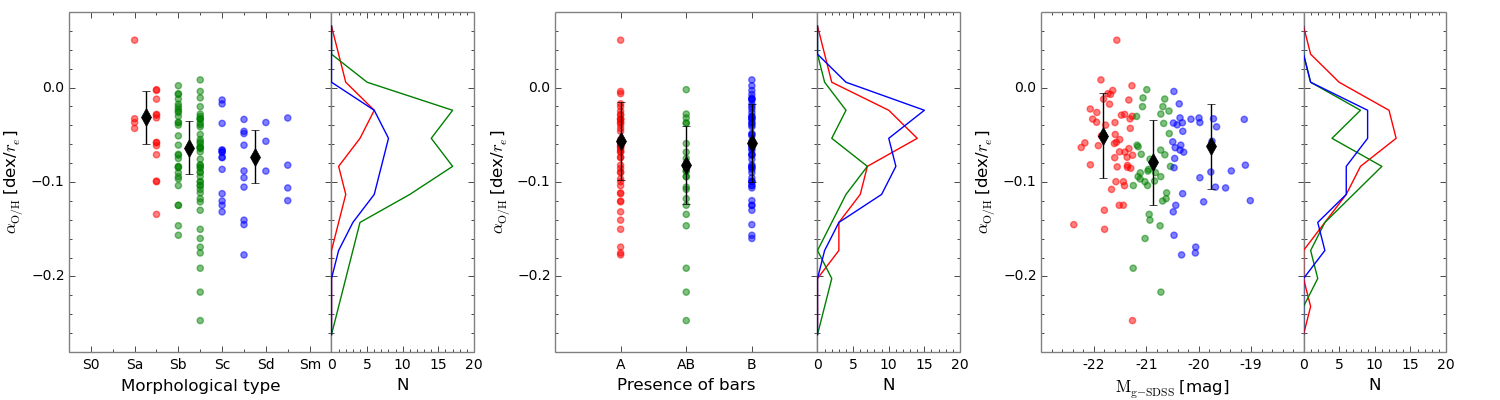}}
\caption{Distribution of the abundance slopes as a function of the morphological type of the galaxies (left panel), depending on the presence or absence of bars (middle panel) and as a function of the $g$-band absolute magnitude of the galaxies (right panel). We also show the histograms for each distribution, where N is
the number of galaxies and colours indicate the different classification types: (i) early spirals, Sa-Sab (red); intermediate spirals, Sb-Sbc (green); late spirals, Sc-Sm (blue) for the left panel. (ii) Clearly unbarred galaxies (red); clearly barred (blue); an intermediate stage (green) for the middle panel. (iii) Luminous galaxies, $M_{g-{\rm SDSS}} < -21.25$ mag (red); intermediate galaxies, $-21.25 < M_{g-{\rm SDSS}} < -20.5$ mag (green); faint galaxies, $M_{g-{\rm SDSS}} > -20.5$ mag (blue) for the right panel. Black diamonds represent the median values for the distributions, together with the standard deviation shown as error bars.}
\label{fig:hists_type}
\end{figure*}

After we derived the oxygen abundance gradient for all galaxies, we analysed whether there was a dependence of the slope on different properties of the galaxies. We focused this analysis on three properties: the differences in the morphological type, the effect of the bars, and the luminosity of the galaxies. We showed in Sect.~\ref{sec:hist} that the oxygen abundance distribution obtained for both the spaxels and \hii\, regions are equivalent and accordingly lead to the same results, therefore we carried this analysis out for the spaxel-wise information alone.

Figure~\ref{fig:hists_type} shows the slope distribution as a function of the morphological type of the galaxy (left panel), the presence or absence of bars (middle panel), and the $g$-band absolute magnitude of the galaxies (right panel). We also show the histograms for each distribution, where  N is the number of galaxies. The limits used in the classification based on the luminosity of the galaxies were chosen to ensure a similar number of elements in each bin (i.e. comparable from a statistical point of view). We tried to follow a similar criterion in the separation by morphological types, but the deficiency of Sa-Sab galaxies and the criterion of not considering Sb as early type prevented us from obtaining a comparable number of objects in each bin.

The slope distribution by morphological types seems to vary in a way that the earlier spirals present flatter gradients than the later type ones. The median values for the distributions together with the standard deviations are
\begin{gather*}
{\rm Sa-Sab:}\:\, \alpha_{O/H} = -0.04 \,\rm{dex}/r_e \text{ and } \sigma = 0.04 \,\rm{dex}/r_e \,(n_{gal} = 17)\\
{\rm Sb-Sbc:}\,\, \alpha_{O/H} = -0.07 \,\rm{dex}/r_e \text{ and } \sigma = 0.05 \,\rm{dex}/r_e \,(n_{gal} = 74)\\ 
{\rm Sc-Sm:}\;\; \alpha_{O/H} = -0.07 \,\rm{dex}/r_e \text{ and } \sigma = 0.04 \,\rm{dex}/r_e \,(n_{gal} = 31).
\end{gather*}
We performed a two-sample Kolmogorov-Smirnov test (KS-test) to check if the differences found between the distributions are significant. The significance level of the KS-test is $5\%$, meaning that values below this limit come from different distributions. We derived a P-value of $7 \%$ for the distributions with the largest differences (between early and late spirals), a P-value of $10 \%$ for the test comparing the early and intermediate types, and a P-value of $92 \%$ between intermediate and late ones. We also performed an Anderson-Darling test (AD-test), which
is more suitable when the samples comprise only few objects, with a resulting P-value of $8 \%$ for the early-late comparison, a P-value of $16 \%$ for the test comparing the early and intermediate type, and a P-value of $73 \%$ in the intermediate-late case. This clearly shows that the observed differences are negligible. 
 
For the distribution of slopes depending on the presence or absence of bars we defined three different groups: galaxies with no bar (A), galaxies that may have a bar, but where the bar is not clearly visible (AB), and clearly barred galaxies (B). The results are as follows:
\begin{gather*}
{\rm A:}\quad \, \alpha_{O/H} = -0.06 \,\rm{dex}/r_e \text{ and } \sigma = 0.05 \,\rm{dex}/r_e \,(n_{gal} = 46)\\
{\rm AB:}\;\: \alpha_{O/H} = -0.09 \,\rm{dex}/r_e \text{ and } \sigma = 0.06 \,\rm{dex}/r_e \,(n_{gal} = 23)\\ 
{\rm B:}\quad \: \alpha_{O/H} = -0.06 \,\rm{dex}/r_e \text{ and } \sigma = 0.04 \,\rm{dex}/r_e \,(n_{gal} = 53).
\end{gather*}
We found negligible differences for these distributions. This was confirmed by the KS test, which gives a P-value of $52 \%$ for the comparison of the A-AB distributions, a P-value of $50 \%$ for the A-B distributions, and a P-value of $10 \%$ for the AB-B distributions (from the AD-tests we obtain P-values of $39 \%$, $27 \%$ and $5 \%$ for each bin, respectively). 

Finally, to analyse the distribution of slopes depending on the luminosity of the galaxies, we divided the sample again into three groups: luminous (L, $M_{g-{\rm SDSS}} < -21.25$ mag), intermediate (I, $-21.25 < M_{g-{\rm SDSS}} < -20.5$ mag), and faint (F, $M_{g-{\rm SDSS}} > -20.5$ mag) galaxies. We obtained these results:
\begin{gather*}
{\rm L:}\quad \alpha_{O/H} = -0.06 \,\rm{dex}/r_e \text{ and } \sigma = 0.05 \,\rm{dex}/r_e \,(n_{gal} = 49)\\
{\rm I:}\quad \: \alpha_{O/H} = -0.09 \,\rm{dex}/r_e \text{ and } \sigma = 0.05 \,\rm{dex}/r_e \,(n_{gal} = 37)\\ 
{\rm F:}\quad \alpha_{O/H} = -0.07 \,\rm{dex}/r_e \text{ and } \sigma = 0.05 \,\rm{dex}/r_e \,(n_{gal} = 36).
\end{gather*}
The statistical tests yield a P-value of $13 \%$ for the KS-test in the case with the largest differences, that is, when comparing luminous and intermediate galaxies. The remaining KS-tests return a P-value of $67 \%$ for the comparison between luminous and faint galaxies and a P-value of $72 \%$ when analysing the intermediate and faint distributions (a P-value of $19 \%$, $50 \%,$ and  $52 \%,$ respectively, for the AD-tests).

Similar results for all these separations are found when using any scale-length normalisation for the radial distance instead of the disc effective radius, either the $r_{25}$ radius or the physical scale of the galaxy (i.e. the radial distance in kpc).

\subsection{Common abundance gradient}\label{sec:gradient}

The fact that the distribution of the oxygen abundance gradients for all the galaxies in the sample is well fitted by a Gaussian function suggests the existence of a characteristic value for the slope, independent of other properties of the galaxies, as the morphological type or the luminosity. This is true when normalising the radial galactocentric distances to the disc effective radius and limiting the fitted interval between 0.5 and 2.0 $r_e$. Below this range ($r/r_e < 0.5$), a nearly flat distribution or even a drop towards the centre is found for some galaxies of the sample. At larger galactocentric distances ($r/r_e > 2.0$), a flattening can be observed in
the abundance gradient of most of the galaxies.

\begin{figure*}
\resizebox{\hsize}{!}{\includegraphics{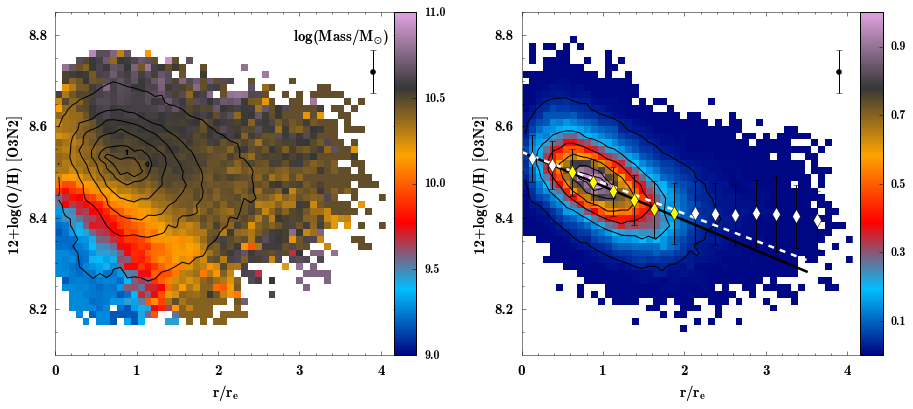}}
\caption{{\it Left:} Radial distribution of the oxygen abundance derived using the O3N2 indicator and the calibration proposed by M13 for all the galaxies in our sample. The black contours show the density distribution of the star-forming spaxels, the outermost one including 95\% of the total number of spaxels and decreasing 20\% in each consecutive contour. The colour bar displays the average stellar mass of each galaxy (in log scale) corresponding to each abundance and radial distance. The average error of the derived oxygen abundances is indicated as an error bar located at the top right side of the panel. For clarity, only the oxygen abundance values with a contribution of at least 1\% of the total number of spaxels are plotted. {\it Right:} Radial density distribution of the oxygen abundance after rescaling the oxygen abundances of each galaxy following the $\mathcal{M}-\mathcal{Z}$ relation derived in \citet{sanchez2013}. As in the left panel, the outermost contour encircles 95\% of the total number of spaxels, decreasing 20\% in each consecutive contour. The diamonds represent the mean oxygen abundance values, with the error bars indicating the corresponding standard deviations, for bins of 0.25 $r_e$. The solid-black line represents the error-weighted linear fit derived for those mean values within the range between 0.5 and 2.0 $r_e$ (yellow diamonds), and the dashed white line represents the linear relation corresponding to the characteristic values of the zero-points and slopes derived in Sect.~\ref{sec:hist} for the individual galaxies assuming a Gaussian distribution for both parameters.}
\label{fig:common_grad}
\end{figure*}

It is easier to illustrate this result if we represent the radial distribution of the oxygen abundance for all the galaxies in the same figure. This is shown in the left panel of Fig.~\ref{fig:common_grad}. The black contours represent the density distribution of the star-forming spaxels. Although the radial gradient can be discerned through the contour plot, the wide range of abundances blurs the result. To clarify the origin of this widening, we colour-code each represented abundance according to the integrated stellar mass of the host galaxy in log scale. For clarity, only the oxygen abundance values with a contribution of at least 1\% of the total number of spaxels are plotted. Adopting this scheme, it is evident that the abundances present a common radial gradient, but with an offset depending on the mass, as expected from the $\mathcal{M}-\mathcal{Z}$ relation (and as discussed in Sect.~\ref{sec:hist}). To remove this dependence on the mass from the map, we rescaled the oxygen abundances of each galaxy following the analytical form of the $\mathcal{M}-\mathcal{Z}$ relation \citep{tremonti2004} derived by \citet{sanchez2013}, applying an offset according to the integrated stellar mass of the galaxy and normalising to the average value for the whole sample of $\sim 8.5$. The outcome is shown in the right panel of Fig.~\ref{fig:common_grad}, where the common abundance gradient presented by all galaxies in the sample can be clearly seen. The abundance distribution is represented again as a density map, both with a colour-code image (normalised to one) and a contour map. The diamonds represent the mean oxygen abundance values, with the error bars indicating the corresponding standard deviations, for bins of 0.25 $r_e$. An error-weighted linear regression (solid black line) to the mean values restricted to the spatial range between 0.5 and 2.0 $r_e$ (yellow diamonds) derives a slope of $\alpha_{O/H} = -\,0.075\,\rm{dex}/r_e$ and $\sigma = 0.016\,\rm{dex}/r_e$, totally compatible with the characteristic slope of $\alpha_{O/H} = -\,0.07\pm0.05 \,\rm{dex}/r_e$ derived in Sect.~\ref{sec:hist} for the individual galaxies assuming a Gaussian distribution (dashed-white line). 

\begin{figure}
\resizebox{\hsize}{!}{\includegraphics{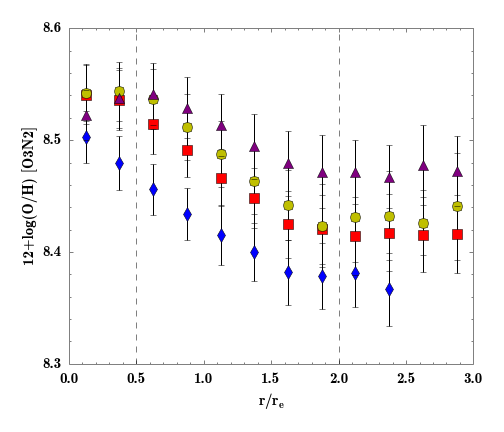}}
\caption{Mean oxygen abundance radial profiles derived considering galaxies in four different bins according to their integrated stellar mass. The limits of the bins were chosen to ensure a similar number of elements in each bin: log (M/M$_{\odot}$) $\leq$ 10.2, blue diamonds; 10.2 < log (M/M$_{\odot}$) $\leq$ 10.5, red squares; 10.5 < log (M/M$_{\odot}$) $\leq$ 10.75, yellow dots; log (M/M$_{\odot}$) $\geq$ 10.75, purple triangles. The symbols represent the mean oxygen abundance values, with the error bars indicating the corresponding standard deviations, for bins of 0.25 $r_e$. Dashed vertical lines delimit the three different behaviours in the oxygen abundance profiles (i.e. drop in the inner parts, common gradient between 0.5 and 2.0 $r_e$, and the flattening in the outer parts).}
\label{fig:grads_mass_bins}
\end{figure}

This characteristic slope is independent of the integrated stellar mass of the galaxies. Figure~\ref{fig:grads_mass_bins} shows the mean radial profiles of the oxygen abundance for our sample of galaxies split in four mass bins with a similar number of elements (log (M/M$_{\odot}$) $\leq$ 10.2, blue diamonds; 10.2 < log (M/M$_{\odot}$) $\leq$ 10.5, red squares; 10.5 < log (M/M$_{\odot}$) $\leq$ 10.75, yellow dots; log (M/M$_{\odot}$) $\geq$ 10.75, purple triangles). The derived gradients (computed between 0.5 and 2.0 $r_e$) of those four profiles are consistent with the characteristic slope displayed by the entire sample.

All these gradients were derived up to two disc effective radii. As we have already mentioned, previous studies have found that galaxies present a flattening in the abundance when covering regions beyond $r_{25}$, observed both in the stellar populations \citep[e.g.][]{yong2006,carraro2007,vlajic2009,vlajic2011} and in the gas \citep[e.g.][]{martin1995, vilchez1996, roy1997, vanzee1998, bresolin2009, rosalesortega2011, bresolin2012, marino2012, lopezsanchez2015, marino2015}. The right panel of Fig.~\ref{fig:common_grad} shows the appearance of the mentioned flattening in the oxygen abundance distribution beyond $\sim $ two effective radii, with around $8\,230$ spaxels at these outer regions. By inspecting the galaxies individually, we detected this flattening in 57 of them, corresponding to 82\% of the galaxies with reliable oxygen abundance values at these large galactocentric distances. The onset of the flattening is always located around 2 $r_e$ even if we separate the sample in different bins according to the integrated stellar mass (see Fig.~\ref{fig:grads_mass_bins}). In the same figure we show that the abundance value of the flattening depends on the mass of the galaxies because of the mentioned $\mathcal{M}-\mathcal{Z}$ relation.

In addition to this flattening in the outer parts, we found that 27 (22\%) galaxies of the sample display some anomalies in their oxygen abundance profiles in the inner parts, namely a nearly flat distribution (8 galaxies, 6.5\%) or even a drop towards the centre (15 galaxies, 15.5\%). This feature has also been found in previous works \citep[e.g.][]{belley1992, rosalesortega2011, sanchez2012b, sanchez2014}. The presence of this feature is not visible when representing the radial distribution of the oxygen abundance for all the galaxies in our sample (Fig.~\ref{fig:common_grad}), but it does appear when separating the sample into different stellar mass bins (Fig.~\ref{fig:grads_mass_bins}), only in the case of the more massive galaxies. While the lowest stellar mass galaxies do not display any sign of this feature, this drop is progressively more evident with increasing galaxy mass.

\section{Discussion and conclusions}\label{sec:discussion}

At the beginning of this article we have pointed out the importance of studying the chemical abundance to understand the evolution of galaxies. In particular, observational studies of the spatial distribution of chemical abundances and its cosmological evolution allow us to set strong constraints for chemical evolution models that try to explain the formation processes of disc galaxies \citep{koeppen1994, edmunds1995, tsujimoto1995, molla1997, prantzos2000, chiappini2001, molla2005, fu2009, pilkington2012, mott2013}. 

The existence of a radial decrease in the chemical abundances of nearby spiral galaxies has been well established by observations for decades \citep{searle1971,comte1975,smith1975,peimbert1979,shaver1983}. Since then, this gradient has been confirmed by other works, restricted to individual galaxies or to limited samples of galaxies \citep{martin1992, kennicutt2003, rosalesortega2011, bresolin2012, marino2012, patterson2012}. With the advent of IFS surveys like CALIFA, abundance studies using larger samples of \hii\, regions have become feasible \citep{sanchez2012b,sanchez2014}, with the same results as in previous studies.

In this work we went a step further and analysed for the first time the oxygen abundance distribution for a large sample of galaxies spaxel by spaxel, taking advantage of the full \mbox{2D} information provided by the CALIFA survey, which improves the statistics. This final sample provides more than 185$\,$000 oxygen abundance values and more than 7$\,$100 \hii\, regions with which to compare our results. The analysis using the \hii\, regions was previously performed by S14 with a different sample of galaxies that were also extracted from the CALIFA mother sample. With spaxel-by-spaxel analysis, we here took approximately four times more line emitting spaxels into account that are associated with star formation regions than in the classical procedure of detecting \hii\, regions. Our results are therefore mainly based on an independent set of measurements, reinforcing and expanding on the S14 results. A more complete 2D study of the oxygen abundance distribution, analysing possible azimuthal variations, will be the topic of a forthcoming work. In this paper, we focused on the radial distribution of the oxygen abundances.

\subsection{Common abundance gradient}\label{sec:discussion1}

Our results confirmed the radially decreasing abundance that was described above and the existence of a characteristic gradient in the oxygen abundance within 0.5-2.0 disc effective radii, independent of other galaxy properties, which is similar to the
result reported previously by \citet{sanchez2012b,sanchez2014}.

The distribution of the derived oxygen abundance gradients is fully compatible with being Gaussian, presenting a characteristic value for the slope of -0.07 dex/$r_e$ and a standard deviation of 0.05 dex/$r_e$. We estimated that at least a 50\% of the dispersion could be justified by the nominal errors in the derivation of the slopes. 

To assess a possible dependence of the gradient distribution on different properties of the galaxies, we studied the abundance gradient distribution of different subsamples according to the morphological type, the effect of bars, and the luminosity of the galaxies. Based on this analysis, we did not find statistically significant differences (in terms of KS-tests) between them.

The results of this analysis contradict some previous studies, which did find a relation between the slope in the gas abundance gradient and some properties of the galaxies, such as the morphology, the mass, or the presence of bars. The dependence of the slope on the morphology of the galaxies is still an open question. Early studies found a correlation between the abundance slope and the morphological type of galaxies, with later types showing steeper gradients \citep{vilacostas1992,oey1993}. Other studies, however, suggested gradients independent of galaxy type \citep{diaz1989,zaritsky1994}, when normalising to a physical scale of the disc (e.g. the $r_{25}$ radius or the disc scale-length $r_d$). The contradictory results might be due to the small and heterogeneous samples and inconsistent methods of measuring metallicity gradients.

For the effect of bars the conclusions likewise disagree. It is well known that roughly $30-40\%$ of the spiral galaxies have a strong bar in optical wavelengths, $60\%$ if we take into account weaker bars \citep[e.g.][]{sellwood1993,marinova2007,sheth2008,aguerri2009,masters2011}. Gas kinematic data show the presence of strong non-circular motions in bars \citep{huntley1978,zurita2004,holmes2015}, indicating that the bar constitutes a major non-axisymmetric component of the galaxy mass distribution \citep{sellwood1993}. Bars have been proposed as a key mechanism in the dynamical evolution of disc galaxies. For instance, they are able to contribute to the redistribution of matter in the galaxy by exchanging angular momentum with the disc, inducing gas flows \citep{lia1992, friedli1998}. This radial movement can produce a mixing and homogenisation of the gas, changing the abundance profiles in the disc and causing a flattening in the gas abundance gradients \citep{friedli1994, friedli1995, portinari2000, cavichia2014}.

Different studies have observed this flattening in the gas abundance gradient of barred galaxies and found a correlation between the abundance slope and the presence of a bar \citep{vilacostas1992,martin1994,zaritsky1994}, independently of the adopted normalisation radius. However, recent works on either stellar or gas-phase metallicity have not found evidence of such a correlation between them \citep{sanchez2012b, sanchez2014, sanchezblazquez2014, cheung2015}. Our results support this absence of a correlation between oxygen gradient and the presence of bars. However, stellar radial migration due to bars may enlarge the effective radius of the disc \citep{debattista2006}, which could compensate for the effect on the oxygen abundance values and produce the same gradient as in absence of bars. We have checked this possibility and confirm that there is no such effect on the disc effective radius, as can be inferred from the bottom right panel of Fig.~\ref{fig:histograms}. This fact, together with the absence of correlation between oxygen gradient and the presence of bar, suggests that bars alone may not have such a strong influence on the chemical evolution of disc galaxies as predicted by simulations. 

We have tried to describe the overall effect of the presence of a bar on the abundance gradients (among other proporties of galaxies), obtaining that bars alone may not affect the chemical evolution of disc galaxies so strongly. However, the relative size of the bar with respect to the disc might play an important role on the derived gradients. To assess this possibility, we need to properly measure the size of the bars, a task that is hampered by the sometimes elusive nature of these structures. This parameter will be determined as a result of the photometric analysis currently carried out by the CALIFA collaboration (M\'endez-Abreu et al., in prep.). Thus, we leave the analysis of this possible effect for a future work.

Finally, no trends with the luminosity or mass of the galaxies have been found in previous studies when normalising to a physical scale of the disc \citep{zaritsky1994, garnett1997, ho2015}; this agrees with our results. We note that several studies that found correlations with some properties of the galaxies measured the abundance gradients in absolute scale (i.e. kpc). When normalising the galactocentric distances to scale lengths such $r_{25}$, $r_d$ or $r_e$, the correlations sometimes disappear. This can partially be understood as a size effect. If galaxies with steeper metallicity gradients measured in dex kpc$^{-1}$ are smaller in their physical sizes (small $r_{25}$), then the steep dex kpc$^{-1}$ metallicity gradients would be compensated for when the galaxy sizes are taken into account. \citet{sanchez2012b,sanchez2013} stated the importance of defining the gradient normalised to the disc effective radius, since this parameter presents a clear correlation with other properties of the galaxies, such as the absolute magnitude, the mass, or the morphological type.

Both the fact that the distribution of the derived oxygen abundance gradients is compatible with a Gaussian distribution and this lack of correlations between the slope values and the analysed properties of galaxies support the existence of a characteristic value for the slope that is common to all type of spiral galaxies (interacting galaxies were not considered in the study and are not included in this statement). 

Several works have detected the mentioned gradient in the MW \citep[e.g.][]{shaver1983, deharveng2000, pilyugin2003, esteban2005, quireza2006, rudolph2006, balser2011}, deriving values for the slope between $-0.04$ and $-0.06$ dex/kpc. If we express the gradient in dex/r$_e$ to easily compare with our results, and considering a value of the disc effective radius for our Galaxy of $r_e=3.6$ kpc \citep{bovy2013}, we obtain a slope between $-0.14$ and $-0.22$ dex/r$_e$, which is slightly larger than the mean value of our slope distribution ($-0.07$ dex/r$_e$, see Table~\ref{tab:grads}). However, deriving the abundance gradient for the MW is not straightforward and presents several obstacles: (i) the measurements are affected by dust and it is difficult to determine distances, which is necessary to derive radial abundance gradients; (ii) the use of different abundance diagnostics such as optical and IR collisionally excited lines (CELs), optical and radio recombination lines (RRL) or thermal continuum emission can yield differences in the results \citep[][]{rudolph2006}; (iii) azimuthal abundance variations have been reported in the MW \citep{pedicelli2009, balser2011} that would complicate any analysis of the radial gradients; and finally (iv) the value of the disc scale-length, and therefore the disc effective radius, is still controversial, with large discrepancies among different results \citep[typical values are in the range $2-4$ kpc, e.g.][]{sackett1997,hammer2007,kruit2011,bovy2013}. All these factors may contribute to an incorrect estimate of the oxygen abundance gradient in our Galaxy, which would explain the differences found in this value between the MW and external galaxies. Another aspect to take into account is the different radial range considered in the derivation of the gradient. While in our work the gradients were computed between 0.5 and 2.0 $r_e$ (see Sect.~\ref{sec:grad}), it is obtained within a typical range between 5 and 15 kpc for the MW, which roughly corresponds to 1.4 and 4.2 $r_e$ \citep[using again a value of 3.6 kpc for the $r_e$,][]{bovy2013}. This fact may also contribute to the differences in the gradient values.

The origin of this negative abundance gradient goes back to the inside-out scenario for the formation of disc galaxies (described in Sect.~\ref{sec:intro}), where the inner parts form first, followed by the formation of the outer regions. Further evidence in the CALIFA survey comes from the analysis of the SP, either from the study of the SFH \citep{perezjimenez2013} or from the study of the radial age gradient \citep[][]{gonzalezdelgado2014, gonzalezdelgado2015, ruizlara2016}. This scenario is understood as a result of the increased timescales of the gas infall with radius and the consequent radial dependence of the SFR. This radial decrease in the chemical abundance has been well established since a long time \citep{searle1971}. However, the statement that this gradient presents a characteristic slope independent of many properties of the galaxies was only recently expressed \citep{sanchez2012b}, which imposes stronger restrictions on our current understanding of disc galaxy growth. The existence of a common gradient in the abundances indicates that the chemical evolution of these disc-dominated galaxies is tightly correlated with the mass growth. 

\subsection{Flattening of the abundance gradient in the outer regions}\label{sec:discussion2}

Several studies analysing the gas content on galaxies have recently found that beyond the isophotal radii $r_{25}$, the metallicity radial distribution flattens to a constant value independent of the galactocentric distance, in contrast to the negative abundance gradient present at smaller distances \citep[e.g.][]{martin1995, vilchez1996, roy1997, vanzee1998, bresolin2009, werk2010, rosalesortega2011, werk2011, bresolin2012, marino2012, sanchez2012b,lopezsanchez2015}. This change in the abundance distribution has not only been detected in the gas, but also in the stellar metallicity \citep[e.g.][]{yong2006,carraro2007,vlajic2009,vlajic2011}. However, all these  works were based on individual galaxies or a very limited sample of objects. S14 represents the first unambiguous detection of such a flattening in a statistically significant large sample of galaxies. This is the first work confirming the flattening by using spaxel-by-spaxel information, increasing the number statistics and improving the spatial coverage across the discs. 

Average radial distributions of oxygen abundances similar to the distribution shown in Fig.~\ref{fig:common_grad} were created for the same subsamples analysed in Sect.~\ref{sec:types} (i.e. according to the morphological type, presence or absence of bars and luminosity). No significant differences were found between them, which means that the flattening at the outer regions seems to be a universal property of spiral galaxies. It is also important to note that although Fig.~\ref{fig:common_grad} was created using the M13 indicator based on the O3N2 index, the flattening is independent of the adopted calibrator; it is present when performing the analysis with all the explored calibrators (as indicated before in Sect.~\ref{sec:calibrators}).

The nature of this flattening is still under debate. Because of the extreme conditions of the outermost parts of galaxies (very low gas densities and long dynamical timescales), these regions play a key role in studying the mechanisms involved in their evolution and, therefore, the existence of this flattening can be of great importance in constraining chemical evolution models.

The flattening, although observed in the more metal poor regions of the galaxies, displays a relatively high abundance value, which ranges between 8.4 and 8.6 (left panel of Fig.~\ref{fig:common_grad}). An estimate of the time necessary to enrich the ISM of the outer parts of discs to these abundance levels, assuming a constant SFR equal to the present observed value, is higher than 10 Gyr \citep[see][for details of the calculation]{bresolin2012}. According to cosmological hydrodynamical simulations, in the inside-out growth scenario the outer regions of galaxy discs are formed during the last $4-6$ Gyr \citep{scannapieco2008,scannapieco2009}. Therefore, if star formation has proceeded at the same rate as observed today, the enrichment of the ISM caused by stellar evolution in these outer parts cannot produce these observed high gas abundance values during the lifetime of these regions.

Different mechanisms have been proposed to explain this enrichment in the outer parts of discs in non-interacting galaxies. One of them is a metal-mixing scenario induced by large-scale processes of angular momentum transport, such as radial gas flows \citep[][among others]{lacey1985, goetz1992, portinari2000, ferguson2001, schonrich2009, bilitewski2012, spitoni2013}, resonance with transient spiral structure \citep{sellwood2002,minchev2012,rovskar2012,daniel2015}, or the overlap of spiral and bar resonances \citep{minchev2010,minchev2011}. In the chemical evolution model developed for the MW by \citet{cavichia2014}, the presence of the bar induces radial gas flows in the disc whose net effect is to produce this flattening of the oxygen gradient at the outer disc. However, as pointed out by the authors, the differences are small compared with the model without the bar and probably cannot be distinguished by the observations. Minor mergers and perturbations caused by orbiting satellite galaxies are also suggested to increase the metal content in the external regions \citep{quillen2009,qu2011,bird2012,lopezsanchez2015}. A slow radial dependence of the star formation efficiency (SFE) at large galactocentric distances is another possible explanation to the flattening \citep{bresolin2012,esteban2013}. An alternative interpretation arises from recent cosmological simulations that propose a balance between outflows and inflows (through `wind recycling' accretion) with the intergalactic medium (IGM) as a mechanism governing the gaseous and metal content of galaxies \citep{oppenheimer2008, oppenheimer2010, dave2011, dave2012}. Finally, it may well be that a fraction of the metals are recycled in the halo of the galaxy and do not escape, being mixed with the incoming gas, as mentioned by \citet{veilleux2005}, and producing the observed flattening.

All these mechanisms are not mutually exclusive, and a possible balance between them could be responsible for the actual chemical evolution of extended discs and the flat abundance of the outermost parts. Moreover, the dominant mechanism could be different for structurally different galaxies. However, our current results do not allow disctinguishing between the different mechanisms; additional information is needed to assess this question.

\subsection{Abundance decrease in the inner regions}\label{sec:discussion3}

As a deviation from the monotonic behaviour of the oxygen abundances increasing towards the centre, several studies have found that galaxies sometimes present a nearly flat distribution of abundances or even a drop in the inner regions, at $r/r_e < 0.3-0.5$ \citep{belley1992, rosalesortega2011, sanchez2012b, sanchez2014}. This feature has also been detected in simulations \citep{molla2005, cavichia2014}.

For our Galaxy, the studies on gas abundance have not been able to properly trace the oxygen abundance at these innermost parts of the disc and therefore were not able to find evidence of this behaviour. However, a chemical evolution model developed by \citet{cavichia2014} has detected this `drop' (or flattening) in the gas abundances of the inner regions as caused by the presence of a bar. According to this model, the presence of the bar induces radial flows that increase the SFR at the corotation radius, which also produces an increase in the oxygen abundance. This results in an apparent decrease or flattening of the abundance in the central regions. A recent study focused on stellar populations \citep{hayden2014} has detected a flattening in the metallicity of the inner regions of the Galaxy, more significant in the low-[$\alpha$/M] ($\alpha$ element abundances) stars. They also explained this feature by the existence of a central bar that produces a mixing of stars in these inner regions.

The analysis presented here can help us to confirm this scenario where the inner drop is caused by the influence of a bar. If the bar were the dominant effect that produces this decrease at inner regions, then it would be expected to detect it more frequently in barred galaxies. However, our analysis yields that only 30\% of galaxies in the sample showing the drop are barred (52\% if we also consider the galaxies that are supposed to have a bar but where it is not clearly visible).

Another explanation proposed by observational studies on gas abundance for external galaxies, like NGC~628 \citep{sanchez2011} and later using CALIFA data in a large sample of galaxies (S14), is that this inner decrease was associated with the presence of a circumnuclear star-forming ring of ionised gas related to the inner Limbland resonance.

To explore the possibility that the abundance drop is due to the presence of a star-forming ring, a visual inspection of the H$\alpha$ intensity maps for the galaxies with signs of this drop was carried out. This analysis showed that 48\% of the galaxies present evidence of a star-forming ring that is spatially located at these inner galactocentric distances ($r/r_e \sim 0.3-0.5$). However, this scenario must be confirmed by a more robust and detailed analysis of the stellar and gas kinematics.

Despite these two possible explanations (influence of a bar or a circumnuclear star-forming ring), the physical origin of this drop is not still well established. On the other hand, if the origin of this drop were due to radial motions of the gas, it would also have an impact on the overall distribution of abundances at larger radii. To explore this possible effect, we derived the mean value for the slope of the abundance gradients only for the galaxies presenting this feature, obtaining a mean $\alpha_{O/H} = -\,0.10\,\rm{dex}/r_e$ and $\sigma = 0.04\,\rm{dex}/r_e$. This value is slightly higher than the one derived for the galaxies without evidence of this inner drop in the abundances, which is $\alpha_{O/H} = -\,0.06\,\rm{dex}/r_e$ and $\sigma = 0.05\,\rm{dex}/r_e$. A KS test suggested that the two distributions are different (P-value of 3\%), consistent with a radial movement of the gas as the cause of this feature. 

Finally, $37\%$ of the galaxies displaying this inner abundance drop neither showed evidence of a star-forming ring nor of a bar. This suggests that another mechanism related to radial movements is responsible for causing this feature. 

To shed some further light on this question, we investigated the presence of this feature depending on the integrated stellar mass of the galaxies. We conclude that the galaxies displaying the strongest oxygen abundance inner drop are the most massive ones (Fig.~\ref{fig:grads_mass_bins}), suggesting that stellar mass plays a key role in shaping the inner abundance profiles.

\vspace{1cm}
In summary, this is the first study that analysed the oxygen abundance distribution for a sample of face-on spiral galaxies spaxel by spaxel. Our final sample of 122 galaxies provided more than 185$\,$000 oxygen abundance values ($\sim$~$8\,200$ of them beyond two disc effective radii) to carry out the analysis. The results were compared to those obtained following the classical procedure of detecting \hii\,regions, leading to equivalent results that point to the same conclusions: (i) the existence of a common abundance gradient, independent of other properties of galaxies, in particular the presence of bars, which seems not to have the flattening effect predicted by numerical simulations; (ii) the existence of a flattening of the abundance gradient in the outer regions of discs, which seems to be a common property of disc galaxies; and (iii) the existence of a drop of the abundance in the inner regions of disc galaxies, only visible in the most massive ones and most probably associated to radial movements of the gas (sometimes a bar or a circumnuclear star-forming ring). These results support the scenario in which disc galaxies present an overall inside-out growth. However, clear deviations were shown with respect to this simple scenario that affect the abundance profiles in both the innermost and outermost regions of galaxies.

The agreement between the two methods arises because the spaxel size of the CALIFA datacubes is of the order of the size of a typical \hii\, region. However, this is not expected for IFS data with better spatial resolution and a smaller spaxel size, where different areas of the \hii\,regions can be resolved and even abundance gradients can be found. Different procedures that can accomodate data that are capable of resolving \hii\,regions may be needed.

\vspace{0.5cm}
\begin{acknowledgements}
This study makes use of the data provided by the Calar Alto Legacy Integral Field Area (CALIFA) survey (\url{http://califa.caha.es/}) based on observations collected at the Centro Astron\'omico Hispano Alem\'an (CAHA) at Calar Alto, operated jointly by the Max-Planck-Institut f\"ur Astronomie and the Instituto de Astrof\'isica de Andaluc\'ia (CSIC).\\

CALIFA is the first legacy survey being performed at Calar Alto. The CALIFA collaboration would like to thank the IAA-CSIC and MPIA-MPG as major partners of the observatory, and CAHA itself, for the unique access to telescope time and support in manpower and infrastructures. The CALIFA collaboration also thanks the CAHA staff for the dedication to this project.\\
We would like to thank the anonymous referee for comments that helped to improve the presentation of our results.\\

We acknowledge financial support from the Spanish {\em Ministerio de Econom\'ia y Competitividad (MINECO)} via grant AYA2012-31935, and from the `Junta de Andaluc\'ia' local government through the FQM-108 project. We also acknowledge support to the ConaCyt funding program 180125. YA acknowledges finantial support from the \emph{Ram\'{o}n y Cajal} programme (RyC-2011-09461). YA and AID acknowledge support from the project AYA2013-47742-C4-3-P from the Spanish MINECO, as well as the `Study of Emission-Line Galaxies with Integral-Field Spectroscopy' (SELGIFS) programme, funded by the EU (FP7-PEOPLE-2013-IRSES-612701). Support for LG is provided by the Ministry of Economy, Development, and Tourism's Millennium Science Initiative through grant IC120009, awarded to The Millennium Institute of Astrophysics, MAS. LG acknowledges support by CONICYT through FONDECYT grant 3140566. RMGD acknowledges support from the Spanish grant AYA2014-57490-P, and from the `Junta de Andaluc\'ia' P12-FQM2828 project. RAM thanks the Spanish program of International Campus of Excellence Moncloa (CEI). IM and AdO acknowledge support from the Spanish MINECO grant AYA2013-42227P. JMA acknowledges support from the European Research Council Starting Grant (SEDmorph, P.I. V. Wild). Support for MM has been provided by DGICYT grant AYA2013-47742-C4-4-P. PSB acknowledges support from the \emph{Ram\'{o}n y Cajal} programme, grant ATA2010-21322-C03-02 from the Spanish MINECO. CJW acknowledges support through the Marie Curie Career Grant Integration 303912.\\

This research makes use of python (\url{http://www.python.org}), of Matplotlib \citep[][]{hunter2007}, a suite of open-source python modules that provides a framework for creating scientific plots, and Astropy, a community-developed core Python package for Astronomy \citep[][]{astropy2013}.

\end{acknowledgements}

\bibliographystyle{aa} 
\bibliography{bibliography}

\clearpage
\onecolumn

\appendix
\section{Fundamental parameters and oxygen abundance information}\label{sec:appendix1} 

In this section we present a table with general information and derived oxygen abundance information for all the galaxies in the sample. From left to right the columns correspond to

\begin{enumerate}
\item the galaxy name,
\item the morphological type, 
\item $g-$band absolute magnitude,
\item the logarithm of the integrated stellar mass in units of solar masses,
\item the disc effective radius ($r_e$, in kpc),
\item the oxygen abundance value at one effective radius,
\item the slope of the oxygen abundance gradient measured between 0.5 and 2.0 $r_e$, and
\item the correlation coefficient of the linear fit (c.c.).
\end{enumerate}

\centering
\begin{longtab}
\tabcolsep=0.45cm
\LTcapwidth=\textwidth
\begin{longtable}{llccccccc}
\caption{Fundamental properties and oxygen abundance information. Details are given in Appendix~\ref{sec:appendix1} above.}\\
\hline\hline\\
Name & Morph & $M_{g}$ & $\log_{10}$ Mass & $r_e$ & [O/H]$_{r_e}$ & $\alpha_{O/H}$ & c.c. \\[0.1cm]
  & type & [mag] & [M$_{\odot}$] & [kpc] & [dex] & [dex/$r_e$] & \\[0.2cm]
{\tiny(a)} & {\tiny(b)} & {\tiny(c)} & {\tiny(d)} & {\tiny(e)} & {\tiny(f)} & {\tiny(g)} & {\tiny(h)}\\[0.1cm]
\hline\\
\endfirsthead
\caption{Continued.}\\
\hline\hline\\
Name & Morph & $M_{g}$ & $\log_{10}$ Mass & $r_e$ & [O/H]$_{r_e}$ & $\alpha_{O/H}$ & c.c. \\[0.1cm]
  & type & [mag] & [M$_{\odot}$] & [kpc] & [dex] & [dex/$r_e$] & \\[0.2cm]
{\tiny(a)} & {\tiny(b)} & {\tiny(c)} & {\tiny(d)} & {\tiny(e)} & {\tiny(f)} & {\tiny(g)} & {\tiny(h)}\\[0.1cm]
\hline\\
\endhead
IC~0159  &  SBdm  &  -19.99  &  9.82  &  5.42  &  8.40  $\pm$  0.06  &  -0.03  $\pm$  0.03  &  0.664 \\
IC~0674  &  SBab  &  -21.43  &  10.91  &  11.55  &  8.44  $\pm$  0.07  &  -0.100  $\pm$  0.015  &  0.871 \\
IC~0776  &  Sdm  &  -19.11  &  9.27  &  8.38  &  8.27  $\pm$  0.05  &  -0.08  $\pm$  0.04  &  0.855 \\
IC~1256  &  SABb  &  -20.73  &  10.37  &  6.82  &  8.58  $\pm$  0.06  &  -0.146  $\pm$  0.015  &  0.983 \\
IC~1683  &  SABb  &  -20.45  &  10.51  &  4.78  &  8.57  $\pm$  0.05  &  -0.08  $\pm$  0.05  &  0.972 \\
IC~4566  &  SBb  &  -21.27  &  10.94  &  10.30  &  8.54  $\pm$  0.04  &  0.00  $\pm$  0.04  &  0.007 \\
MCG-01-10-019  &  SABbc  &  -20.62  &  10.21  &  12.04  &  8.40  $\pm$  0.06  &  -0.12  $\pm$  0.04  &  0.925 \\
NGC~0001  &  Sbc  &  -21.07  &  10.80  &  5.99  &  8.56  $\pm$  0.06  &  -0.020  $\pm$  0.014  &  0.470 \\
NGC~0036  &  SBb  &  -21.77  &  10.90  &  14.57  &  8.53  $\pm$  0.07  &  -0.04  $\pm$  0.03  &  0.890 \\
NGC~0160  &  Sa  &  -21.55  &  11.06  &  11.44  &  8.53  $\pm$  0.08  &  0.05  $\pm$  0.07  &  0.851 \\
NGC~0165  &  SBb  &  -21.09  &  10.59  &  13.04  &  8.51  $\pm$  0.04  &  -0.07  $\pm$  0.04  &  0.966 \\
NGC~0171  &  SBb  &  -21.26  &  10.72  &  6.98  &  8.58  $\pm$  0.06  &  -0.072  $\pm$  0.016  &  0.894 \\
NGC~0214  &  SABbc  &  -21.60  &  10.85  &  6.66  &  8.58  $\pm$  0.05  &  -0.059  $\pm$  0.012  &  0.991 \\
NGC~0234  &  SABc  &  -21.40  &  10.65  &  6.36  &  8.58  $\pm$  0.05  &  -0.07  $\pm$  0.02  &  0.982 \\
NGC~0237  &  SBc  &  -20.72  &  10.28  &  4.27  &  8.56  $\pm$  0.05  &  -0.066  $\pm$  0.011  &  0.977 \\
NGC~0257  &  Sc  &  -21.60  &  10.80  &  8.67  &  8.57  $\pm$  0.05  &  -0.074  $\pm$  0.014  &  0.913 \\
NGC~0309  &  SBcd  &  -22.38  &  10.83  &  13.37  &  8.50  $\pm$  0.05  &  -0.15  $\pm$  0.03  &  0.986 \\
NGC~0477  &  SABbc  &  -21.24  &  10.50  &  14.26  &  8.43  $\pm$  0.08  &  -0.19  $\pm$  0.06  &  0.990 \\
NGC~0496  &  Scd  &  -20.93  &  10.35  &  10.86  &  8.46  $\pm$  0.06  &  -0.14  $\pm$  0.03  &  0.970 \\
NGC~0570  &  SBb  &  -21.24  &  10.96  &  8.61  &  8.47  $\pm$  0.03  &  -0.10  $\pm$  0.03  &  0.965 \\
NGC~0716  &  SABb  &  -20.72  &  10.58  &  5.54  &  8.50  $\pm$  0.06  &  -0.09  $\pm$  0.02  &  0.971 \\
NGC~0768  &  SBc  &  -21.33  &  10.55  &  10.40  &  8.41  $\pm$  0.08  &  -0.013  $\pm$  0.014  &  0.469 \\
NGC~0776  &  SBb  &  -21.26  &  10.70  &  6.88  &  8.56  $\pm$  0.06  &  -0.030  $\pm$  0.013  &  0.822 \\
NGC~0787  &  Sa  &  -21.30  &  10.96  &  6.68  &  8.50  $\pm$  0.06  &  -0.04  $\pm$  0.02  &  0.965 \\
NGC~0873  &  Scd  &  -21.17  &  10.41  &  4.32  &  8.53  $\pm$  0.03  &  -0.061  $\pm$  0.011  &  0.995 \\
NGC~0941  &  Scd  &  -19.13  &  9.35  &  2.87  &  8.40  $\pm$  0.06  &  -0.034  $\pm$  0.019  &  0.794 \\
NGC~0976  &  Sbc  &  -21.50  &  10.80  &  5.72  &  8.48  $\pm$  0.00  &  -0.068  $\pm$  0.018  &  0.753 \\
NGC~0991  &  SABcd  &  -19.38  &  9.61  &  3.65  &  8.41  $\pm$  0.06  &  -0.09  $\pm$  0.02  &  0.851 \\
NGC~1070  &  Sb  &  -21.73  &  11.00  &  7.08  &  8.53  $\pm$  0.05  &  -0.006  $\pm$  0.014  &  0.470 \\
NGC~1093  &  SBbc  &  -20.82  &  10.75  &  7.27  &  8.52  $\pm$  0.06  &  -0.090  $\pm$  0.015  &  0.979 \\
NGC~1094  &  SABb  &  -21.55  &  10.72  &  7.45  &  8.50  $\pm$  0.05  &  -0.084  $\pm$  0.012  &  0.950 \\
NGC~1659  &  SABbc  &  -21.15  &  10.59  &  6.35  &  8.50  $\pm$  0.06  &  -0.094  $\pm$  0.012  &  0.956 \\
NGC~1667  &  SBbc  &  -22.02  &  10.80  &  4.91  &  8.56  $\pm$  0.04  &  -0.033  $\pm$  0.011  &  0.936 \\
NGC~2253  &  SBbc  &  -21.06  &  10.47  &  2.53  &  8.57  $\pm$  0.06  &  -0.011  $\pm$  0.012  &  0.558 \\
NGC~2347  &  SABbc  &  -21.66  &  10.72  &  6.31  &  8.53  $\pm$  0.06  &  -0.108  $\pm$  0.012  &  0.998 \\
NGC~2449  &  SABab  &  -20.98  &  10.83  &  5.85  &  8.56  $\pm$  0.05  &  -0.002  $\pm$  0.011  &  0.252 \\
NGC~2486  &  SBab  &  -20.64  &  10.58  &  9.55  &  8.49  $\pm$  0.05  &  -0.07  $\pm$  0.03  &  0.989 \\
NGC~2487  &  SBb  &  -21.68  &  10.77  &  12.04  &  8.54  $\pm$  0.05  &  -0.007  $\pm$  0.015  &  0.078 \\
NGC~2530  &  SABd  &  -20.96  &  10.19  &  9.78  &  8.40  $\pm$  0.06  &  -0.09  $\pm$  0.03  &  0.994 \\
NGC~2540  &  SBbc  &  -21.15  &  10.50  &  8.28  &  8.49  $\pm$  0.06  &  -0.063  $\pm$  0.013  &  0.753 \\
NGC~2604  &  SBd  &  -19.97  &  9.64  &  4.09  &  8.35  $\pm$  0.06  &  -0.037  $\pm$  0.013  &  0.216 \\
NGC~2639  &  Sa  &  -21.59  &  11.18  &  4.88  &  8.57  $\pm$  0.06  &  -0.037  $\pm$  0.019  &  0.723 \\
NGC~2730  &  SBcd  &  -20.56  &  10.11  &  7.83  &  8.46  $\pm$  0.05  &  -0.049  $\pm$  0.014  &  0.945 \\
NGC~2805  &  Sc  &  -20.30  &  10.03  &  7.86  &  8.44  $\pm$  0.04  &  -0.11  $\pm$  0.03  &  0.991 \\
NGC~2906  &  Sbc  &  -20.14  &  10.39  &  3.41  &  8.60  $\pm$  0.04  &  -0.034  $\pm$  0.011  &  0.929 \\
NGC~2916  &  Sbc  &  -21.57  &  10.75  &  8.08  &  8.56  $\pm$  0.04  &  -0.100  $\pm$  0.013  &  0.986 \\
NGC~3057  &  SBdm  &  -19.01  &  9.18  &  6.02  &  8.27  $\pm$  0.08  &  -0.12  $\pm$  0.07  &  0.997 \\
NGC~3106  &  Sab  &  -22.16  &  11.31  &  13.31  &  8.50  $\pm$  0.05  &  -0.06  $\pm$  0.05  &  0.922 \\
NGC~3381  &  SBd  &  -19.74  &  9.68  &  3.07  &  8.50  $\pm$  0.03  &  -0.057  $\pm$  0.012  &  0.972 \\
NGC~3614  &  SABbc  &  -20.72  &  10.21  &  9.56  &  8.42  $\pm$  0.06  &  -0.22  $\pm$  0.05  &  0.999 \\
NGC~3687  &  SBb  &  -20.47  &  10.28  &  4.72  &  8.50  $\pm$  0.06  &  -0.156  $\pm$  0.010  &  0.989 \\
NGC~3811  &  SBbc  &  -20.92  &  10.45  &  5.39  &  8.57  $\pm$  0.04  &  -0.076  $\pm$  0.010  &  0.923 \\
NGC~4047  &  Sbc  &  -21.41  &  10.68  &  5.05  &  8.57  $\pm$  0.05  &  -0.104  $\pm$  0.012  &  0.972 \\
NGC~4185  &  SABbc  &  -21.35  &  10.72  &  10.31  &  8.55  $\pm$  0.05  &  -0.082  $\pm$  0.012  &  0.898 \\
NGC~4210  &  SBb  &  -20.46  &  10.28  &  4.42  &  8.58  $\pm$  0.04  &  -0.085  $\pm$  0.011  &  0.935 \\
NGC~4470  &  Sc  &  -20.37  &  9.98  &  3.46  &  8.42  $\pm$  0.02  &  -0.017  $\pm$  0.008  &  0.037 \\
NGC~4961  &  SBcd  &  -19.97  &  9.68  &  3.87  &  8.42  $\pm$  0.05  &  -0.095  $\pm$  0.012  &  0.973 \\
NGC~5000  &  SBbc  &  -21.32  &  10.66  &  8.16  &  8.55  $\pm$  0.06  &  -0.065  $\pm$  0.012  &  0.855 \\
NGC~5016  &  Sbc  &  -20.61  &  10.23  &  4.64  &  8.54  $\pm$  0.05  &  -0.112  $\pm$  0.013  &  0.974 \\
NGC~5056  &  SABc  &  -21.43  &  10.45  &  8.65  &  8.44  $\pm$  0.05  &  -0.125  $\pm$  0.010  &  0.981 \\
NGC~5157  &  SBab  &  -21.81  &  11.27  &  10.06  &  8.55  $\pm$  0.07  &  -0.01  $\pm$  0.02  &  0.442 \\
NGC~5205  &  SBbc  &  -19.66  &  9.88  &  3.79  &  8.52  $\pm$  0.06  &  -0.04  $\pm$  0.02  &  0.690 \\
NGC~5320  &  SABbc  &  -20.72  &  10.29  &  8.17  &  8.50  $\pm$  0.07  &  -0.028  $\pm$  0.018  &  0.985 \\
NGC~5376  &  SABb  &  -20.31  &  10.47  &  3.76  &  8.60  $\pm$  0.05  &  -0.037  $\pm$  0.013  &  0.685 \\
NGC~5378  &  SBb  &  -20.64  &  10.62  &  7.36  &  8.53  $\pm$  0.04  &  -0.01  $\pm$  0.03  &  0.314 \\
NGC~5406  &  SBb  &  -21.94  &  11.19  &  9.72  &  8.53  $\pm$  0.06  &  -0.03  $\pm$  0.02  &  0.863 \\
NGC~5480  &  Scd  &  -20.30  &  10.14  &  3.13  &  8.57  $\pm$  0.04  &  -0.046  $\pm$  0.012  &  0.989 \\
NGC~5520  &  Sbc  &  -19.77  &  9.85  &  3.39  &  8.49  $\pm$  0.04  &  -0.089  $\pm$  0.009  &  0.982 \\
NGC~5533  &  Sab  &  -21.91  &  11.23  &  14.16  &  8.49  $\pm$  0.05  &  -0.06  $\pm$  0.03  &  0.811 \\
NGC~5622  &  Sbc  &  -20.41  &  10.22  &  6.54  &  8.53  $\pm$  0.08  &  -0.039  $\pm$  0.014  &  0.813 \\
NGC~5633  &  Sbc  &  -20.44  &  10.27  &  2.50  &  8.60  $\pm$  0.03  &  -0.064  $\pm$  0.007  &  0.917 \\
NGC~5656  &  Sb  &  -21.11  &  10.62  &  4.28  &  8.55  $\pm$  0.05  &  -0.097  $\pm$  0.011  &  0.984 \\
NGC~5657  &  SBbc  &  -20.48  &  10.28  &  6.95  &  8.47  $\pm$  0.07  &  -0.038  $\pm$  0.012  &  0.757 \\
NGC~5665  &  SABc  &  -20.66  &  10.22  &  4.72  &  8.50  $\pm$  0.05  &  -0.04  $\pm$  0.03  &  0.904 \\
NGC~5720  &  SBbc  &  -21.79  &  10.85  &  12.32  &  8.53  $\pm$  0.05  &  -0.130  $\pm$  0.017  &  0.920 \\
NGC~5732  &  Sbc  &  -20.06  &  9.93  &  6.52  &  8.47  $\pm$  0.06  &  -0.18  $\pm$  0.03  &  0.999 \\
NGC~5735  &  SBbc  &  -21.02  &  10.38  &  8.05  &  8.49  $\pm$  0.06  &  -0.16  $\pm$  0.02  &  0.997 \\
NGC~5772  &  Sab  &  -21.56  &  10.94  &  9.39  &  8.52  $\pm$  0.06  &  -0.06  $\pm$  0.02  &  0.995 \\
NGC~5829  &  Sc  &  -21.36  &  10.53  &  10.77  &  8.43  $\pm$  0.05  &  -0.074  $\pm$  0.014  &  0.927 \\
NGC~5888  &  SBb  &  -22.06  &  11.29  &  11.45  &  8.54  $\pm$  0.06  &  -0.02  $\pm$  0.02  &  0.690 \\
NGC~5947  &  SBbc  &  -21.02  &  10.37  &  7.70  &  8.49  $\pm$  0.05  &  -0.100  $\pm$  0.013  &  0.983 \\
NGC~5957  &  SBb  &  -20.34  &  10.28  &  4.69  &  8.56  $\pm$  0.05  &  -0.06  $\pm$  0.02  &  0.958 \\
NGC~6004  &  SBbc  &  -21.30  &  10.71  &  7.96  &  8.53  $\pm$  0.06  &  -0.033  $\pm$  0.020  &  0.784 \\
NGC~6063  &  Sbc  &  -20.05  &  10.06  &  6.34  &  8.48  $\pm$  0.05  &  -0.17  $\pm$  0.02  &  0.986 \\
NGC~6154  &  SBab  &  -21.63  &  10.95  &  9.42  &  8.52  $\pm$  0.07  &  -0.003  $\pm$  0.019  &  0.028 \\
NGC~6155  &  Sc  &  -20.28  &  10.12  &  3.47  &  8.55  $\pm$  0.03  &  -0.068  $\pm$  0.009  &  0.973 \\
NGC~6301  &  Sbc  &  -22.23  &  11.02  &  15.71  &  8.51  $\pm$  0.04  &  -0.063  $\pm$  0.014  &  0.879 \\
NGC~6941  &  SBb  &  -21.85  &  10.93  &  12.10  &  8.54  $\pm$  0.06  &  -0.05  $\pm$  0.03  &  0.719 \\
NGC~7321  &  SBbc  &  -22.06  &  10.81  &  9.58  &  8.52  $\pm$  0.05  &  -0.082  $\pm$  0.011  &  0.964 \\
NGC~7364  &  Sab  &  -21.41  &  10.89  &  5.92  &  8.57  $\pm$  0.07  &  -0.028  $\pm$  0.013  &  0.777 \\
NGC~7466  &  Sbc  &  -21.29  &  10.83  &  12.32  &  8.47  $\pm$  0.05  &  -0.085  $\pm$  0.018  &  0.933 \\
NGC~7489  &  Sbc  &  -21.79  &  10.40  &  10.59  &  8.41  $\pm$  0.06  &  -0.150  $\pm$  0.013  &  0.970 \\
NGC~7591  &  SBbc  &  -21.35  &  10.75  &  8.43  &  8.52  $\pm$  0.07  &  -0.084  $\pm$  0.020  &  0.960 \\
NGC~7625  &  Sa  &  -19.71  &  10.07  &  2.00  &  8.54  $\pm$  0.04  &  -0.033  $\pm$  0.011  &  0.889 \\
NGC~7653  &  Sb  &  -21.11  &  10.50  &  6.17  &  8.52  $\pm$  0.05  &  -0.090  $\pm$  0.015  &  0.995 \\
NGC~7691  &  SBbc  &  -20.97  &  10.21  &  8.94  &  8.45  $\pm$  0.07  &  -0.10  $\pm$  0.04  &  0.967 \\
NGC~7716  &  Sb  &  -20.56  &  10.32  &  4.46  &  8.49  $\pm$  0.04  &  -0.025  $\pm$  0.010  &  0.850 \\
NGC~7782  &  Sb  &  -21.94  &  11.18  &  10.82  &  8.54  $\pm$  0.06  &  -0.037  $\pm$  0.018  &  0.857 \\
NGC~7819  &  Sc  &  -20.68  &  10.39  &  8.36  &  8.49  $\pm$  0.06  &  -0.12  $\pm$  0.02  &  0.985 \\
NGC~7824  &  Sab  &  -21.47  &  11.30  &  11.60  &  8.48  $\pm$  0.05  &  -0.03  $\pm$  0.05  &  0.991 \\
UGC~00005  &  Sbc  &  -21.50  &  10.83  &  9.56  &  8.52  $\pm$  0.04  &  -0.055  $\pm$  0.009  &  0.987 \\
UGC~00036  &  SABab  &  -20.94  &  10.98  &  8.64  &  8.51  $\pm$  0.06  &  -0.13  $\pm$  0.04  &  0.997 \\
UGC~01918  &  SBb  &  -20.53  &  10.56  &  7.95  &  8.53  $\pm$  0.05  &  -0.08  $\pm$  0.04  &  0.996 \\
UGC~02311  &  SBbc  &  -21.59  &  10.74  &  7.95  &  8.52  $\pm$  0.04  &  -0.050  $\pm$  0.010  &  0.942 \\
UGC~03253  &  SBb  &  -20.43  &  10.68  &  6.54  &  8.52  $\pm$  0.04  &  -0.125  $\pm$  0.012  &  0.938 \\
UGC~03973  &  SBbc  &  -21.86  &  10.74  &  7.66  &  8.50  $\pm$  0.05  &  0.008  $\pm$  0.009  &  0.209 \\
UGC~04195  &  SBb  &  -20.68  &  10.50  &  8.02  &  8.52  $\pm$  0.07  &  -0.020  $\pm$  0.013  &  0.789 \\
UGC~04262  &  SABbc  &  -21.26  &  10.60  &  12.31  &  8.49  $\pm$  0.10  &  -0.25  $\pm$  0.03  &  0.953 \\
UGC~04308  &  SBc  &  -20.95  &  10.30  &  6.73  &  8.52  $\pm$  0.06  &  -0.087  $\pm$  0.011  &  0.964 \\
UGC~04375  &  Sbc  &  -19.90  &  10.16  &  6.12  &  8.49  $\pm$  0.04  &  -0.121  $\pm$  0.015  &  0.948 \\
UGC~05108  &  SBb  &  -21.51  &  10.89  &  17.80  &  8.49  $\pm$  0.04  &  -0.12  $\pm$  0.05  &  0.977 \\
UGC~07012  &  SABcd  &  -19.68  &  9.45  &  5.28  &  8.40  $\pm$  0.08  &  -0.11  $\pm$  0.03  &  0.995 \\
UGC~08781  &  SBb  &  -21.69  &  11.05  &  17.57  &  8.48  $\pm$  0.06  &  -0.02  $\pm$  0.03  &  0.904 \\
UGC~09291  &  Scd  &  -20.32  &  10.34  &  7.64  &  8.44  $\pm$  0.06  &  -0.18  $\pm$  0.03  &  0.986 \\
UGC~09476  &  Sbc  &  -20.49  &  10.21  &  5.89  &  8.49  $\pm$  0.05  &  -0.057  $\pm$  0.012  &  0.971 \\
UGC~09777  &  Sbc  &  -20.47  &  10.31  &  7.67  &  8.48  $\pm$  0.08  &  -0.00  $\pm$  0.06  &  0.266 \\
UGC~09842  &  SBbc  &  -21.18  &  10.63  &  14.27  &  8.50  $\pm$  0.06  &  -0.031  $\pm$  0.015  &  0.533 \\
UGC~11649  &  SBab  &  -20.84  &  10.66  &  6.23  &  8.53  $\pm$  0.07  &  -0.10  $\pm$  0.03  &  0.890 \\
UGC~12224  &  Sc  &  -20.48  &  9.97  &  7.82  &  8.47  $\pm$  0.06  &  -0.13  $\pm$  0.05  &  0.987 \\
UGC~12633  &  SABab  &  -20.35  &  10.37  &  7.09  &  8.47  $\pm$  0.00  &  -0.03  $\pm$  0.02  &  0.666 \\
UGC~12816  &  Sc  &  -20.35  &  9.82  &  8.56  &  8.42  $\pm$  0.06  &  -0.07  $\pm$  0.02  &  0.767 \\
UGC~A021  &  SBdm  &  -19.48  &  9.45  &  5.12  &  8.35  $\pm$  0.06  &  -0.11  $\pm$  0.06  &  0.700 \\
\end{longtable}
\end{longtab}
\newpage

\clearpage
\twocolumn

\end{document}